\documentclass[numberedappendix,twocolumn,twocolappendix]{openjournal}
\usepackage{color}
\usepackage{verbatim}
\usepackage{ulem}
\usepackage{graphicx}
\usepackage{dcolumn}
\usepackage{bm}
\usepackage[hypertexnames=false]{hyperref}
\usepackage{tabularx}
\usepackage{amsmath}
\usepackage{numprint}
\usepackage{booktabs}
\usepackage{xspace}

\usepackage{natbib}
\begin{document}
\title{Emulator-Based Inference of Cosmological Subgrid Models}
\author{Nesar Ramachandra$^{1}$, Nicholas Frontiere$^{1}$, Michael Buehlmann$^{1}$, Kelly R. Moran$^{2}$, J.D.~Emberson$^{1}$,\\ Katrin Heitmann$^{3}$, Salman Habib$^{1,3}$}
\affiliation{$^{1}$Computational Science Division, Argonne National Laboratory, Lemont, IL 60439, USA}
\affiliation{$^{2}$Statistical Sciences Group, CCS Division, Los Alamos National Laboratory, Los Alamos, NM 87545, USA}
\affiliation{$^{3}$High Energy Physics Division, Argonne National Laboratory, Lemont, IL 60439, USA}

\begin{abstract}

The formation of structure in the Universe at large scales is dominated by gravity, with baryonic physics becoming significant at $\sim{\rm Mpc}$ scales. To capture the impact of baryonic physics, cosmological simulations must model gas dynamics and a host of relevant astrophysical processes. A recent extension of the Hardware/Hybrid Accelerated Cosmology Code (HACC) couples its gravity solver with a modern smoothed particle hydrodynamics method. This extension incorporates sub-resolution models for chemical enrichment, black hole and star formation, AGN kinetic and thermal feedback, supernova-driven feedback, galactic winds, and metal-line cooling. We present an inference framework based on high-fidelity emulators to aid in model calibration against observational targets, e.g., the galaxy stellar mass function, radial gas density profiles, and the cluster gas fraction. The emulators are trained on simulation suites comprising 64 boxes with side-length $128\,h^{-1}$Mpc and 16 boxes with side-length $256\,h^{-1}$Mpc with $2\times 512^3$ and $2\times 1024^3$ particles, respectively. Our analysis reveals two distinct AGN kinetic feedback modes -- a low-feedback mode yielding strong agreement with the observed radial gas density profiles of massive X-ray clusters, and a high-feedback mode providing a better fit to cluster gas fraction data, but systematically underestimating gas densities in inner regions. 

\end{abstract}
\maketitle
\section{Introduction}

Simulating the detailed evolution of structure formation in the Universe is a complex task involving significant trade-offs between computational costs and physical accuracy. The simplest case involves gravity-only (GO) simulations, where matter interacts solely via self-consistent gravitational interactions in an expanding universe. The GO approaches enable the simulation of cosmological volumes at high resolution up to $\sim$Gpc scales (e.g., \citealt{potter2017pkdgrav3, heitmann2019outer, ishiyama2021uchuu,2021ApJS..252...19H, frontiere2022farpoint}). Because the computational demands are not overwhelming, this approach is commonly used for exploring cosmological parameter spaces via simulation suites, statistical surrogate models, and Bayesian inference schemes (e.g., \citealt{Heitmann2006}, \citealt{Habib2007}, \citealt{Heitmann:2013bra, DeRose_2019}). The simplifications inherent to GO simulations, however, preclude them from directly producing baryonic observables, such as gas densities or galaxy colors. To address this drawback, heuristic models of the galaxy-halo connection are employed in post-processing of the simulation results (e.g., \citealt{2009ApJ...696..620C, Mead_2015, wechsler2018connection}). 
While such approaches offer substantial utility, they are subject to obvious limitations in terms of properly capturing a variety of dynamically complex astrophysical and gasdynamic effects. 

Cosmological hydrodynamic simulations treat idealized gas dynamics along with a set of small-scale baryonic processes such as radiative cooling, star and black hole formation, galactic winds, and feedback from active galactic nuclei (AGN) (For recent reviews, see \citealt{Crain2023}, \citealt{Vogelsberger2020}, \citealt{Valentini2025}.) These simulations allow for a more detailed and physical representation of baryonic effects but require empirical sub-resolution physics models (or `subgrid' models) to account for processes occurring at scales much smaller than their resolution in both space and time. Assumptions and simplifications that underlie subgrid models are often responsible for differences between the results of cosmological hydro simulations, rather than differences in the solvers; conversely, consistent results for some observables can result even with different subgrid modeling assumptions.

Cosmological hydrodynamic simulations are computationally intensive, which imposes trade-offs between simulated volume, resolution, and the number of runs that can be performed. Several projects have prioritized single large-volume runs using fixed subgrid models – for example, MillenniumTNG \citep{Pakmor2023MillenniumTNG}, the TNG300 run of IllustrisTNG \citep{nelson2019}, SIMBA \citep{dave2019}, and BAHAMAS \citep{Pfeifer2020bahamas}. These flagship simulations yield rich datasets and broadly reproduce many observed galaxy properties, calibrated with a particular subgrid tuning and thus show varying degrees of agreement with different observables. Conversely, other efforts have focused on suites of numerous smaller simulations that systematically vary subgrid parameters to explore and constrain the uncertain physics. For example, the EAGLE project involved a series of 50 Mpc simulations in which feedback parameter values were adjusted to match the $z\sim0$ galaxy stellar mass function ($\rm{GSMF}$) observations \citep{crain2015eagle, schaye2015eagle}. The CAMELS initiative likewise consists of thousands of $\sim(25~\mathrm{Mpc})^3$ simulations designed to span wide ranges of cosmological parameters and subgrid feedback parameters \citep{Villaescusa-Navarro2021camels}. More recently, the DREAMS project has introduced thousands of simulations that simultaneously vary dark matter microphysics, galaxy formation physics, and cosmology to systematically probe the astrophysical effects \citep{rose2025introducing, rose2025dreams}. Notably, some modern projects combine both approaches: the FLAMINGO \citep{kugel2023flamingo, schaye2023flamingo} and COLIBRE \citep{schaye2025colibre, chaikin2025colibre} collaborations first ran a dedicated ensemble of simulations to tune their stellar and AGN feedback models against key observables, and then applied those calibrated models in large production simulations. This calibrated-suite strategy enables a more rigorous exploration of subgrid parameter space and uncertainty, complementing the insights gained from the traditional single-run mega-simulations. 

With the advent of GPU-accelerated computational resources, it is now possible to speed up detailed hydrodynamic simulations quite substantially, by roughly an order of magnitude. Consequently, systematic parameter space exploration, even in these more complex scenarios, has become possible, enabling controlled studies of how subgrid effects interact. As the model space grows more complex, such systematic approaches become essential -- `by-hand' or manual tuning will become impractical, and is, in any case, scientifically unattractive. Another point is that simulation ensembles with significantly larger volumes for individual runs are now possible; this is a crucial issue when considering group and cluster-scale observables, which may otherwise be statistics-limited.

The ability to carry out large ensembles of simulations is also essential in the construction of fast surrogate models. These models -- built using summary statistics from (primarily GO) simulation suites -- have multiple uses in cosmological studies. In particular, emulation techniques based on machine learning melded with statistical methods have been developed, beginning with frameworks like the CosmicEmu for the matter power spectrum~\citep{2009ApJ...705..156H,2010ApJ...715..104H,2010ApJ...713.1322L}, with subsequent extensions to smaller scales, broader redshift coverage, and larger parameter spaces~\citep{2014ApJ...780..111H,2017ApJ...847...50L}. Emulators for other summary statistics, such as the halo mass function~\citep{bocquet2020miratitan}, galaxy power spectra~\citep{Kwan2015, Wibking_2019}, and the concentration-mass relation~\citep{Kwan_2013} have also been constructed; the utility of this approach is further demonstrated in the results from large emulation projects such as in~\cite{Nishimichi_2019,kobayashi2020accurate, Kwan2023, Moran2023}.

Another, less accurate, approach predating the use of emulators is based on the use of fitting functions motivated, e.g., by halo models (e.g., Halofit by \citealt{smith2003stable, takahashi2012revising}). Since extensive hydrodynamic simulation suites were not available earlier, the first inclusion of baryonic effects was carried out in the context of halo models (e.g., \citealt{Mead_2015}). To go beyond these initial efforts, it is first essential to recognize key distinctions between GO and hydrodynamic simulations equipped with subgrid models, as emulation and inference in these two cases possess distinctly different characteristics. 

Emulators based on GO simulations are typically designed to explore and infer cosmological parameters, leveraging well-defined physics that governs the evolution of matter on large scales and relatively smooth parametric dependencies. In contrast, emulators for hydrodynamic simulations address the calibration of heuristic subgrid models that encapsulate unresolved baryonic processes. This introduces an additional layer of subjectivity, as the subgrid models and associated parameters are not rooted in first principles but are instead empirical approximations (albeit with a hopefully physical basis). In addition, subgrid models are inherently stochastic, resolution-dependent, and highly sensitive to the local environment, further complicating the inference process required for model calibration. Moreover, the errors and systematics associated with observational data and their interpretation, such as biases in stellar mass estimates or measurements of the gas fraction, are tightly coupled to the calibration of subgrid parameters. 

Another key difference between the GO scenario and hydrodynamic simulations is the computational expense: survey-scale hydrodynamic runs are so resource-intensive that they have only recently become practical, enabled by access to large-scale computing resources. Considering the cost of a single large hydrodynamic simulation, it is critically important to have robustly calibrated subgrid models. Accurate emulators for a number of observational summary statistics provide an important toolkit to calibrate subgrid model parameters against different sets of observations, each of which may have their own specific set of circumstances (e.g., sky coverage, systematics, different instrumental modeling, and other effects). Of course, different choices of calibration variables will result in different values of subgrid parameters and likely different predicted outcomes for observationally relevant variables not used in the calibration process. A systematic assessment of how calibration choices propagate into inferred subgrid parameters and downstream predictions is therefore essential; emulators are natural and efficient tools to enable such analyses.

Several large-scale hydrodynamic simulations have calibrated feedback efficiency using observed stellar mass estimates either from the $\rm{GSMF}$ or the stellar-to-halo mass relation (SHMR). The black hole mass to stellar mass relationship ($\rm{BHMSM}$) is another statistic commonly employed to calibrate feedback efficiencies linked to black hole growth. However, calibration on just these quantities alone does not ensure agreement with all galaxy properties, often requiring complementary target observations. For instance, the EAGLE simulations \citep{crain2015eagle} incorporated the sizes of disk galaxies, allowing for a more faithful reproduction of key galaxy scaling relations. IllustrisTNG \citep{nelson2019} extended this approach by also considering the cosmic star formation rate density and halo gas fractions. In contrast, SIMBA \citep{dave2019} tuned the feedback efficiencies solely against the $\rm{GSMF}$ and the $\rm{BHMSM}$ relations, calibrating the latter through accretion efficiency rather than AGN feedback efficiency. Horizon-AGN \citep{2017MNRAS.467.4739K} inferred the stellar feedback efficiency from the Starburst spectrophotometric model \citep{1999ApJS..123....3L}. Their AGN feedback model, similar to EAGLE and IllustrisTNG, was tuned with respect to the $\rm{BHMSM}$ relation. While parameter choices have traditionally resulted from meticulous manual tuning involving numerous trials, given the complexity of the problem, scaling across multiple parameters and simulation conditions is best achieved via an automated approach. 

In this paper, we develop a suite of Gaussian Process-based emulators to estimate multiple observable summary statistics and employ the emulators within a joint calibration framework. We discuss the construction of the emulators, comprising experimental design, training data synthesis, and statistical techniques to perform interpolation across subgrid modeling parameters. We then use the emulators to infer subgrid model parameters under different conditions of joint likelihoods, simulation fidelity, and bias factors. Additionally, we develop emulators for observables not used in the calibration, and discuss the impact of baryonic physics on these quantities. A related approach was recently employed for the FLAMINGO simulations \citep{kugel2023flamingo}, where Gaussian process emulators were used to calibrate subgrid parameters against the $\rm{GSMF}$ and cluster gas fraction $f_\mathrm{gas}$ values. We develop a complementary framework for CRK-HACC \citep{frontiere2023simulating}, which was used to obtain the calibrated subgrid parameters employed in the Frontier-E simulation \citep{frontiere2025cosmological}, a four trillion particle run with a $(4.655\,\rm{Gpc})^3$ volume. We include the cluster gas density $\mathrm{CGD}$ profiles as an additional calibration target compared to \citet{kugel2023flamingo}, which reveals distinct AGN kinetic feedback regimes not captured by calibrations to integrated gas fractions.

The structure of the paper is as follows: We discuss the subgrid models and the connected subgrid model parameters in our cosmological simulations in \autoref{sec:data}. \autoref{sec:calibration} describes our calibration approach, including the observations as targets. The emulator construction and inference is done in two phases, where the first phase includes more subgrid model parameters and a larger number of simulations, but smaller volumes, while the second phase is restricted to two subgrid model parameters, a smaller number of simulations in larger volumes. We provide results for additional observables that were not used for the calibration process in \autoref{sec:addon}. Finally, related technical details concerning the Gaussian process method are discussed in \autoref{sec:appendix_gp}.

\section{Simulation Framework}
\label{sec:data}

The GPU-accelerated simulation code CRK-HACC (Conservative Reproducing Kernel HACC; \citet{frontiere2023simulating}) is the basis for the cosmological hydrodynamic simulations in this study. CRK-HACC leverages a higher-order SPH method,  Conservative Reproducing Kernel Smoothed Particle Hydrodynamics (CRKSPH;~\citealt{frontiere2017crksph}) to model gas dynamics alongside gravity, also GPU-accelerated. In addition, localized baryonic processes are included via subgrid models to capture sub-resolution astrophysics. CRK-HACC's solvers are designed for high performance and scalability, as are the associated analysis routines, which leverage GPUs and in situ methods. Simulation outputs for a typical CRK-HACC run include detailed dark matter halo and galaxy properties and profiles, which are used to predict multi-wavelength observables. In this section, we summarize the subgrid model implementation and associated parameters. Full details of the subgrid model implementations are presented in \cite{frontiere2025modeling}.

\subsection{Subgrid Models}

The subgrid models implemented in CRK-HACC include prescriptions for radiative and metal-line cooling, star formation, galactic winds, and active galactic nucleus (AGN) feedback. Each model incorporates parameters that can be tuned to match observations, while others are held fixed. The tuning parameters are not only coupled nonlinearly but are also resolution dependent, making accurate emulation beneficial not just for calibration but also for quantifying parameter degeneracies and assessing model fidelity. The subgrid model implementations relevant for the current paper are:

\begin{itemize}

\item \textbf{Radiative cooling and heating}: Radiative processes enable gas to cool and collapse, driving the formation of stars and galaxies. We assume that the gas is optically thin and in photoionization equilibrium, subject to a spatially uniform, time-dependent ultraviolet background (UVB) radiation field, following the rates provided in \cite{faucher2020cosmic}. Metal-line cooling is based on the total metallicity $Z$, similar to the approach of \cite{wiersma2009effect}, with cooling rates computed using the photoionization code CLOUDY\footnote{\url{https://gitlab.nublado.org/cloudy/cloudy/-/wikis/home}}.

\item \textbf{Star formation and chemical enrichment: } A hybrid multiphase approach (originally described in~\citealt{springel2003cosmological}) models the star-forming interstellar medium (ISM) using an effective equation of state, with stars forming stochastically on a prescribed star formation timescale. This module includes chemical enrichment from single stellar populations, informed by the FIRE simulations~\citep{hopkins2018fire}, enabling tracking of metals injected into the ISM by supernovae and stellar winds.

\item \textbf{Kinetic galactic outflow model: } Galactic winds, which regulate star formation and redistribute gas and metals within and beyond galaxies, are modeled following the subgrid prescription of the TNG simulations~\citep{pillepich2018simulating}. We vary the TNG parameters for the wind velocity $\kappa_\text{w}$, which scales with the local one-dimensional dark matter velocity dispersion, and the energy injection parameter $e_\text{w}$ which sets the available energy for outflows. 

\item \textbf{Active galactic nuclei (AGN) model: } Thermal and kinetic feedback from AGN are implemented similarly to the TNG model~\citep{weinberger2016simulating}. Black holes are initialized with a seed mass $M_\text{seed}$ and accrete mass according to the Bondi-Hoyle-Lyttleton accretion rate \citep{hoyle1939,bondi1944}. The accreted mass is continuously converted into thermal energy with a fixed, preset efficiency. When the accretion rate drops below a threshold fraction $\chi$ of the Eddington rate (defined as in the TNG simulation), kinetic feedback is activated. This mode is governed by two parameters: The velocity of the feedback events, $v_\text{kin}$, and the kinetic feedback efficiency, $\epsilon_\text{kin}$.

\end{itemize}

In this paper, we calibrate the five subgrid parameters $\theta_{\rm sub} = $\{$\kappa_\text{w}$, $e_\text{w}$, $M_\text{seed}$, $v_\text{kin}$, $\epsilon_\text{kin}$\} introduced via the subgrid models above to match selected observations and to impose numerical robustness. All other parameters that enter the complete subgrid prescription are held fixed (see \cite{frontiere2025modeling} for the parameter choices). In addition, we fix the $\Lambda$CDM cosmology parameters to $\Omega_{\rm cdm}=0.26067$, 
$\Omega_{\rm b}h^2=0.02242$, 
$H_0=67.66$, 
$\sigma_8=0.8102$, 
$n_s=0.9665$. 
We assume spatial flatness and massless neutrinos.

\section{Subgrid Parameter Calibration}
\label{sec:calibration}

We employ a joint calibration framework to constrain subgrid model parameters through simulation-based emulators coupled with Bayesian inference, following the approach described in, e.g., \cite{Heitmann2006, Habib2007, Higdon2008}. In the following, we first describe the three canonical observables studied for calibration in this paper in \autoref{sec:obs}. 

We then describe our calibration process, which is carried out in two phases. In the first phase (Phase-1), \autoref{sec:phase1}, we construct Gaussian-process emulators for all three observables using 64 CRK-HACC simulations spanning the full five-parameter subgrid space. Phase-1 allows us to successfully constrain three subgrid parameters. In Phase-2, \autoref{sec:phase2}, we systematically study the remaining two of the five subgrid model parameters in more detail, with a focus on one observable. We use results from 16 simulations, each with eight times larger volumes than the individual runs used in Phase-1. This improvement in the statistical sampling of massive halos is needed to provide more information on the remaining two parameters, both of which are associated with the properties of galaxy groups and clusters. The numerical parameters of the Phase-1 and Phase-2 simulations are shown in \autoref{tab:simulations}.

\begin{table*}
\centering
\caption{Numerical parameters in Phase-1 and Phase-2 simulations. The columns list: the simulation campaign, number of simulations in the suite $N_{\textrm{sim}}$, comoving box size $L$, number of particles (dark matter and baryons) $N$, initial simulation redshift $z_\mathrm{in}$, dark matter particle mass $m_\mathrm{DM}$, baryonic particle mass $m_\mathrm{g}$, comoving gravitational softening length $\epsilon_\mathrm{com}$, and gravitational softening length $\epsilon_\mathrm{prop}$. }

\label{tab:simulations}
\begin{tabular}{lcccccccc}
\hline\hline
Campaign & 
$N_{\textrm{sim}}$ &
$L$ & 
$N$ & 
$z_\mathrm{in}$ & $m_\mathrm{DM}$ & $m_\mathrm{g}$ & $\epsilon_\mathrm{com}$ & $\epsilon_\mathrm{prop}$ \\
           & & ($h^{-1}$Mpc) &  &  & ($h^{-1}\mathrm{M}_\odot$) & ($h^{-1}\mathrm{M}_\odot$) & ($h^{-1}$kpc) & ($h^{-1}$kpc) \\
\hline
Phase-1 & 64 & 128 & $2 \times 512^3$ & 200 & $1.13 \times 10^{9}$ & $2.12 \times 10^{8}$ & 10 & 6 \\
Phase-2 & 16 & 256 & $2 \times 1024^3$ & 200 & $1.13 \times 10^{9}$ & $2.12 \times 10^{8}$ & 10 & 6 \\
\hline
\end{tabular}
\end{table*}

\subsection{Observations Used for Calibration}
\label{sec:obs}

Calibration of the subgrid modeling parameters is necessarily related to observables sensitive to baryonic physics. These observables should cover a wide range of halo masses and satisfy basic requirements of statistical coverage and control of measurement uncertainties. Additionally, for full use in cosmological analyses, the dependence on cosmological parameters also enters, which we defer to later work.

The calibration targets considered here are 1) $\rm{GSMF}$, 2) $\mathrm{CGD}$ profile, and 3) $f_\mathrm{gas}$. These observables reflect complementary aspects of the subgrid physics included in the CRK-HACC simulation suite, and their combination helps in jointly reproducing observations with the least bias. The $\rm{GSMF}$ directly constrains certain aspects of galaxy formation physics and galaxy clustering. It has been widely adopted as a primary calibration observable for galaxy formation models in SIMBA \citep{2019MNRAS.486.2827D} and Illustris-TNG \citep{nelson2019}. On the other hand, the $f_\mathrm{gas}$ in groups and clusters focuses on the baryon distribution and feedback processes in massive halos. In the calibration of the FLAMINGO simulation, \cite{kugel2023flamingo} used the $\rm{GSMF}$ and $f_\mathrm{gas}$ jointly for calibration. For stacked measurements, the $\mathrm{CGD}$ profile requires a sufficient sampling of massive clusters. It has, therefore, not been typically used for parameter calibration, due to the limited volumes of most hydrodynamic simulation ensembles. In this work, we introduce the $\mathrm{CGD}$ profile as an additional calibration target, addressing similar regimes as $f_\mathrm{gas}$, but at higher halo masses. The observational data sets used for our calibration along with the curation information from sky surveys are listed below.

\begin{itemize}
    \item \textbf{Galaxy stellar mass functions:} The $\rm{GSMF}$ describes the number density of galaxies as a function of stellar mass. It is a key observable in galaxy evolution studies, helping to constrain models of star formation, feedback processes, and galaxy assembly across cosmic time \citep{baldry2008galaxy, davidzon2017cosmos2015}. We use the data from the Galaxy And Mass Assembly (GAMA, \citealt{Driver2022}) survey that provides high-resolution spectroscopic information, enabling precise measurements of the $\rm{GSMF}$ across various redshifts and environments. We focus on the galaxy stellar masses in the range of $5 \times 10^9 \,$M$_\odot$ to $3 \times 10^{11}\,$M$_\odot$ for $M_\star\,/\,$M$_\odot$. In the simulation, the $\rm{GSMF}$ is measured by counting galaxies in 40 logarithmically distributed stellar-mass bins in the interval $[10^{8.5}, 10^{13}]$\,$M_\odot$.

    \item \textbf{Cluster gas density profiles:} The $\mathrm{CGD}$ observable refers to the radial gas-density profiles within galaxy clusters, often parameterized through models such as the generalized Navarro-Frenk-White (gNFW) profile \citep{1996MNRAS.278..488Z}. These profiles are essential for interpreting X-ray and Sunyaev-Zel’dovich (SZ) observations of clusters \citep{vikhlinin2006chandra}. The correct modeling of the $\mathrm{CGD}$ is also key to reproducing the thermodynamical state of the intracluster medium (ICM) in simulations. In our simulations, we tabulate the $\mathrm{CGD}$ within a cluster radius range in $r/R_{\rm 500c}$ from $0.015$ to $2.75$. ($R_{\rm 500c}$ is the radius within which the density is 500 times the critical density.) In each simulation, the $\mathrm{CGD}$ profiles of all halos with $M_{200c} > 10^{13}$\,$h^{-1}M\odot$ are normalized by the corresponding $R_{500c}$, and stacked by logarithmically interpolating each profile to a common $r/R_{500c}$ spacing. As an observational target, we calibrate against the low-redshift stacked ICM gas-density profile from \citet{mcdonald2017remarkable}, derived from deep \textit{Chandra} observations of 27 massive X-ray--selected clusters at $z \lesssim 0.1$ (drawn from the \citet{vikhlinin2009} sample after a mass cut).

    \item \textbf{Cluster gas fractions:} The ratio of gas mass to total mass in galaxy clusters provides insights into baryonic feedback mechanisms. Observations of $f_\mathrm{gas}$~within $R_{\rm 500c}$ constrain the baryon content of clusters and the efficiency of feedback mechanisms from AGN or supernovae that expel or redistribute gas \citep{ettori2009cluster, pratt2019galaxy}. Accurate modeling of $f_\mathrm{gas}$~also provides insights into the large-scale baryon distribution in the Universe. In our simulations, we tabulate $f_\mathrm{gas}$~in the cluster mass $M_{\rm 500c}$ range between $10^{13.5} \,h^{-1}$M$_\odot$ and $10^{14.3}\,h^{-1}$M$_\odot$. We measure the median gas fraction in 50 logarithmically distributed halo-mass bins between $[10^{12}, 10^{16}]$\,$h^{-1}M_\odot$. 

    For the observational targets, we adopt the dataset compiled in \cite{kugel2023flamingo} (see their Tables 4 and 5), which aggregates cluster gas-fraction measurements and mass proxies across various surveys: The X-ray component spans relaxed nearby Chandra clusters, a large archival Chandra sample $(z > 0.1)$, the representative XMM-Newton REXCESS sample, and additional group- to cluster-scale studies with optical/IR or flux-limited selections \citep{vikhlinin2006chandra, maughan2008images, pratt2010gas, rasmussen2009temperature, sun2009chandra, lin2011baryon, lagana2013comprehensive, gonzalez2013galaxy, sanderson2013baryon, lovisari2015scaling, pearson2017galaxy, lovisari2020x}. This is complemented with the weak gravitational lensing data from \cite{akino2022hsc} (HSC-XXL) and \cite{hoekstra2015canadian}, and LoCuSS WL–multi-wavelength scaling relations for high masses \citep{mulroy2019locuss}. In the simulations, we measure the median gas fraction in 50 logarithmically distributed halo-mass bins between $[10^{12}, 10^{16}]$\,$h^{-1}M_\odot$ to obtain $f_\mathrm{gas}$. 
    
\end{itemize}

One challenge when calibrating subgrid model parameters is that the subgrid effects depend on stochastic interactions influenced by variables such as the local matter density and simulation parameters such as the particle mass. This behavior varies significantly with resolution, and therefore, the calibration is usually only valid for a specific mass resolution set by the combination of the number of particles evolved and the simulation volume. In this paper, we only investigate simulations at a single resolution, motivated by maintaining resolution alignment with the multi-trillion-particle Frontier-E simulation in \cite{frontiere2025cosmological}. Investigations of the effects of resolution on parameter choices will be presented in a forthcoming paper. Another obstacle arises due to the limited statistics of higher-mass halos in smaller simulation volumes. Obtaining a sufficient number of galaxies and clusters in the training simulations is crucial for deriving robust ensemble statistics. In particular, the $\mathrm{CGD}$ observable in the simulations is sensitive to the number of massive clusters, whose under-representation in smaller volumes may result in systematic biases.

\subsection{Phase-1: Initial Subgrid Model Parameter constraints}
\label{sec:phase1}

In this section, we describe Phase-1 of the calibration process in the two-part scenario that was sketched earlier. The simulation suite and inference procedure are designed to cover the five subgrid model parameters $\theta_{\rm sub} = $\{$\kappa_\text{w}$, $e_\text{w}$, $M_\text{seed}$, $v_\text{kin}$, $\epsilon_\text{kin}$\}. We employ a symmetric Latin hypercube design to generate $64$ points within the parameter ranges shown in~\autoref{tab:params1}. The minimum and maximum limits of parameter values in the experimental design were empirically chosen to ensure coverage of the observational targets while providing sufficient dynamic range for posterior estimation.

\begin{table}[ht]
\centering
\begin{tabular}{lcc}
\toprule
\textbf{$\theta_{\text{sub}}$} & \textbf{min($\theta_{\text{sub}}$)} & \textbf{max($\theta_{\text{sub}}$)} \\ 
\midrule
$\kappa_\text{w}$ & 2 & 4 \\
$e_\text{w}$ & 0.2 & 1 \\
$M_\text{seed} [M_\odot h^{-1}]$ & $0.6 \times 10^6$ & $1.2 \times 10^6$ \\
$v_\text{kin} [\rm{km/s}]$ & $0.1 \times 10^4$ & $1.2 \times 10^4$ \\
$\epsilon_\text{kin}$ & $0.2$ & $12$ \\ 
\bottomrule
\end{tabular}
\caption{Priors for subgrid parameters $\theta_{\text{sub}}$ in the experimental design of Phase-1, showing the minimum and maximum values for each subgrid parameter}
\label{tab:params1}
\end{table}

This experimental design ensures an efficient, space-filling sampling of the parameter hypercube. For each point in the parameter space, we run a CRK-HACC hydrodynamic simulation with side-length $L = 128\,h^{-1}$Mpc evolving $N_\textrm{p}=2\times512^3$ particles. The corresponding mass resolution and softening lengths are shown in \autoref{tab:simulations}.  

Halos are identified using a friends-of-friends algorithm (FoF, \citealt{1985ApJ...292..371D, 1993ApJ...416....1K}) applied to all the dark matter particles within a linking length of 0.168 times the average inter-particle spacing. Spherical overdensity (SOD) halo masses $M_\mathrm{200c}$ and $M_{500c}$, with corresponding radii, $R_{200c}$ and $R_{500c}$, are measured around the potential minimum of each FoF halo with a density threshold of $200$ and $500 \rho_\mathrm{crit}$, respectively. 
The gas fraction is measured within $R_{500c}$, and the radial gas-density profiles are constructed in 50 logarithmically distributed bins extending out to $\sim 2 \times R_{\rm 200c}$. 

Galaxies are identified with the DBSCAN clustering algorithm described in \cite{1996kddm.conf..226E} from all stellar particles with a proper linking length of $l=50$\, kpc (in proper units) and a $n_\mathrm{neigh}=10$. We measure the stellar mass in a 3D aperture of $50$\, kpc (in proper units) centered at the potential minimum of each galaxy. More details on the galaxy definition and identification can be found in \cite{frontiere2025modeling}. 

In post-processing, we extract the $\rm{GSMF}$, $f_\mathrm{gas}$, and $\mathrm{CGD}$ for each of the simulations. Mass limits and binning of $\rm{GSMF}$ and $f_\mathrm{gas}$ are mentioned in \autoref{sec:obs}. Objects with SOD Mass $M_{500c} > 10^{14}\, h^{-1}M_\odot$ are selected as massive clusters for $\mathrm{CGD}$ calculations.

\subsubsection{Emulation with Five Subgrid Parameters} 

Given a set of parameters, emulators are designed to provide controlled approximations of summary statistics essentially instantaneously. In this paper, we utilize Gaussian processes (GPs) for the higher-order interpolation tasks (Other approaches include neural networks  \citep{Agarwal_2014} and polynomial chaos \citep{euclidemu2019}). Given the small number of training points $(N_{\textrm{sim}}=64)$, GPs are a particularly effective choice, especially when combined with Principal Component Analysis (PCA) for the dimensionality reduction of the summary statistics \citep{Higdon2008}. This methodology has been applied successfully in numerous cosmological emulators \citep{Heitmann2006, 2007A&A...464..399H, 2010ApJ...715..104H, 2009ApJ...705..156H, 2010ApJ...713.1322L}.

The emulator development in this work is carried out with SEPIA (Simulation-Enabled Prediction, Inference, and Analysis), a Python code developed at Los Alamos National Laboratory \citep{james_gattiker_2020_4048801} that implements the Bayesian emulation and calibration methodology described in \cite{Higdon2008}. For the $\rm{GSMF}$, $f_\mathrm{gas}$, and $\mathrm{CGD}$ at redshift $z=0$, we first create truncated PCA bases that capture 95\% of the total variance of the original datasets. Then, GP interpolation is carried out to map from the parameter space of $\theta_{\rm sub}$ to the space of the truncated PCA weights. (For a detailed explanation of the emulator construction, the reader is referred to \autoref{sec:appendix_gp}.) During the emulator deployment, the saved basis sets are multiplied by the GP predictions, producing the summary statistics at new subgrid parameter values. We note that these trained emulators in the suite are not calibrated against observations and are solely surrogates of the simulation outputs. The mean emulator errors, estimated via leave-one-out cross-validation, are less than 1\% for the $\rm{GSMF}$ (in log-space), approximately 7\% for $f_\mathrm{gas}$, and 9\% for the $\mathrm{CGD}$. The larger errors for cluster-related statistics reflect the intrinsic noise in the training data in the Phase-1 simulation volumes.

\subsubsection{Parameter Sensitivity}

\begin{figure*}[t]
    \centering
    \includegraphics[width=0.32\linewidth]{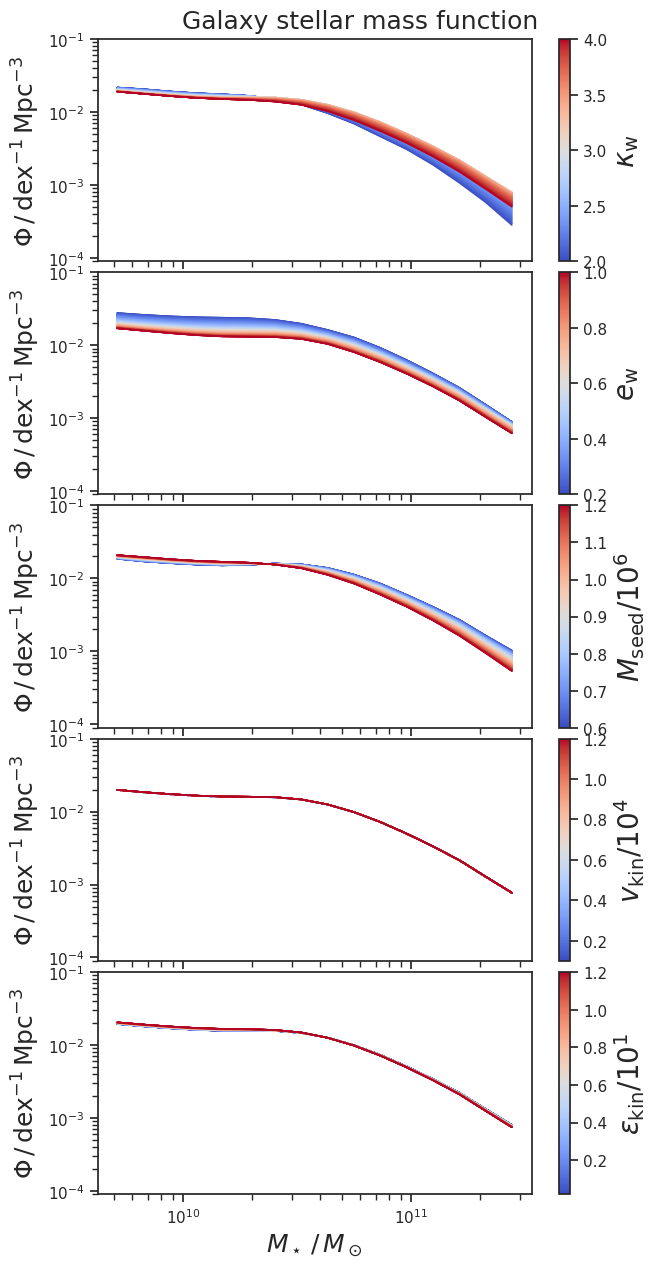}
    \includegraphics[width=0.32\linewidth]{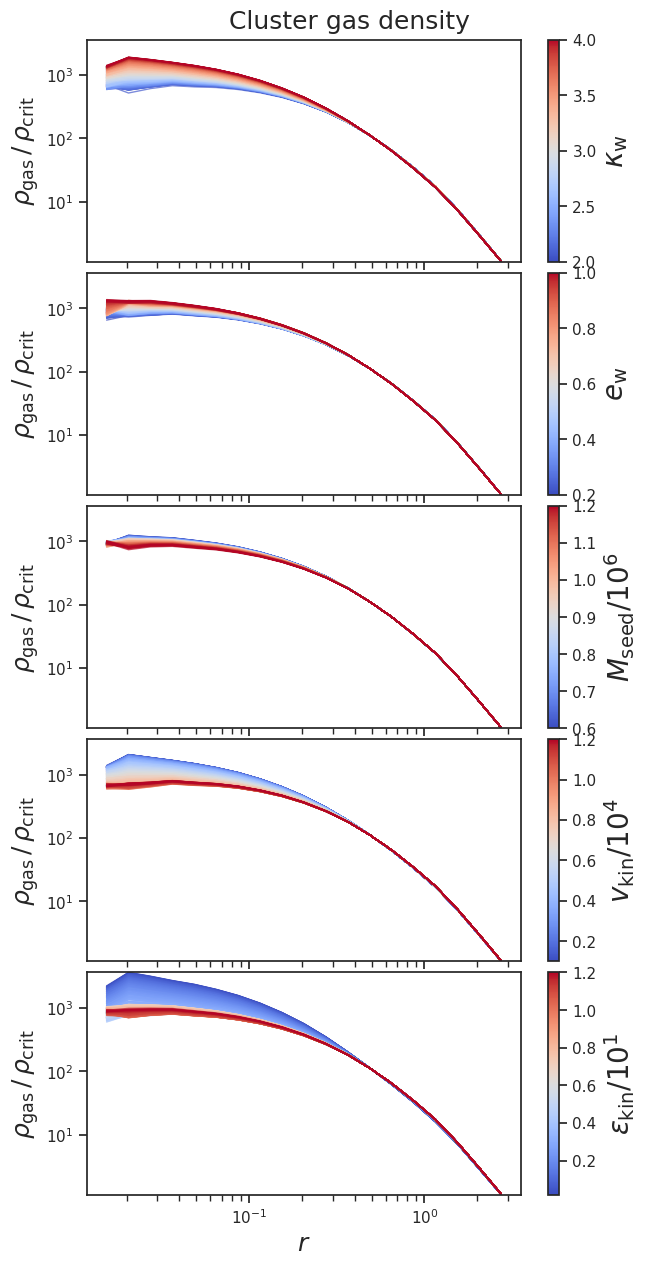}
    \includegraphics[width=0.32\linewidth]{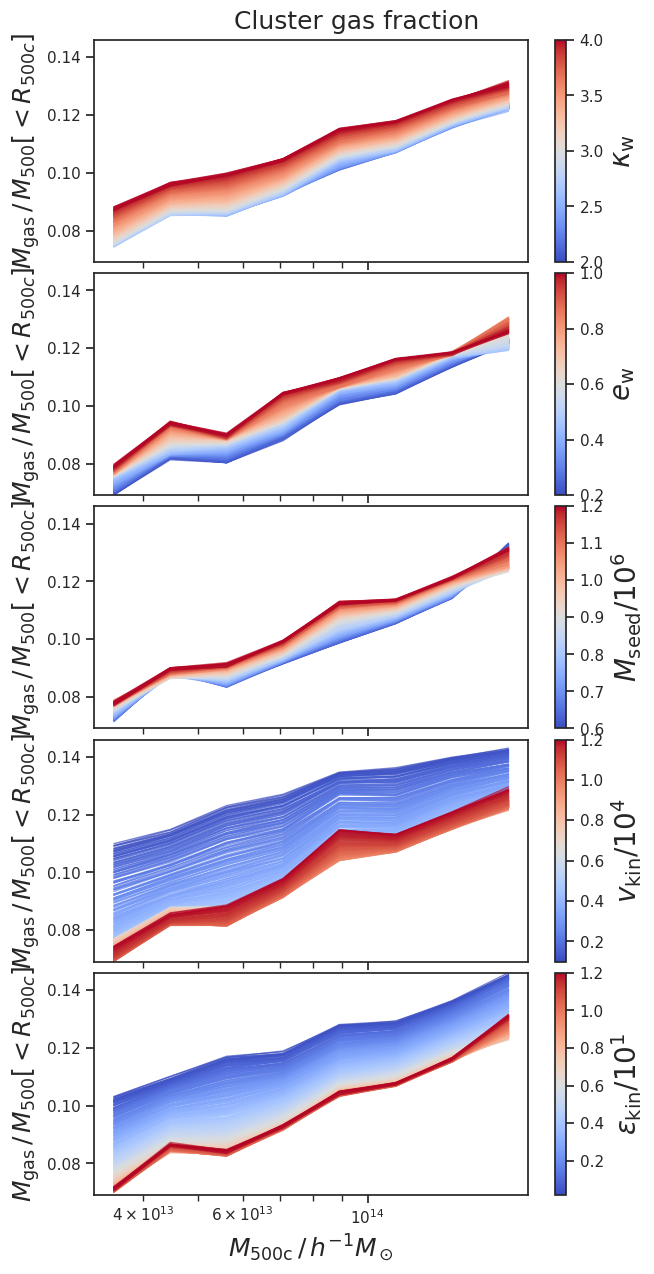}
    \caption{Results for the sensitivity analysis for the three observables ($\rm{GSMF}$, $\mathrm{CGD}$, and $f_\mathrm{gas}$) using emulator predictions. The impact of five subgrid parameters is investigated. Within each panel, the variation of a summary statistic is shown when a single parameter is varied, while the rest are fixed to the center of the experimental design.}
    \label{fig:sensitivity}
\end{figure*}
 
Before calibration, it is useful to investigate the sensitivity of the observables to variations in individual parameters. We investigate the variation of the $\rm{GSMF}$, $f_\mathrm{gas}$, and $\mathrm{CGD}$ under changing input subgrid parameters. To understand the first-order change, we fix the parameters to the center of the Latin hypercube, \{$\kappa_\text{w}=3$, $e_\text{w}=0.6$, $M_\text{seed}/10^6 = 0.9$, $v_\text{kin}/10^4=0.65~\rm{km/s}$, $\epsilon_\text{kin}/10^1=0.61$\}, and allow one parameter at a time to vary. The parameter being varied is scanned across its entire range, as shown in \autoref{fig:sensitivity}. 

The kinetic feedback parameters have a limited effect on the $\rm{GSMF}$, whereas variations in $\kappa_\text{w}$, $e_\text{w}$, and $M_\text{seed}$ have a greater impact on the changes in the $\rm{GSMF}$. In particular, increasing the energy injection parameter from our wind model $e_\text{w}$~leads to a decrease in the number density of galaxies across the entire range of stellar mass. Increasing the wind velocity predominantly leads to an increase in the number of highly massive ($ M_\star > 2 \times 10^{10}\,$M$_\odot$) galaxies. The middle panels of \autoref{fig:sensitivity} show $\mathrm{CGD}$ variations. The inner regions of the radial gas density profiles ($r<0.3 R_{500c}$) are most sensitive to subgrid parameter changes, with the highest variations connected to the kinetic feedback parameters. Finally, $f_\mathrm{gas}$ variations show that the increase in kinetic feedback parameters $v_\text{kin}$~and $\epsilon_\text{kin}$~lowers the total gas content in the galaxy clusters, whereas the other three subgrid parameters show a positive correlation with $f_\mathrm{gas}$.

While this sensitivity analysis provides valuable information about the influence of the parameters on the different observables, the results should be interpreted as local, first-order effects. Interactions across multiple parameters are not captured, and the responses are evaluated only in the vicinity of a single reference point (the center of the Latin hypercube in this case). Consequently, the observed trends do not fully represent the system’s behavior across other regions of the parameter space. 

\subsubsection{Parameter Constraints Using MCMC}\label{sec:mcmc} 

The speed and precision of our emulators enable quick parameter inference using traditional Bayesian inference schemes like Markov Chain Monte Carlo (MCMC) methods. In addition, exploring the posterior distribution allows us to study the covariance between the subgrid parameters that arise from the observational data vectors. For these Bayesian inference runs, we use a Gaussian-like prior distribution for each subgrid parameter $\theta_{\rm sub}$ within a bounded range defined by our experimental design. Outside the bounds, the prior probabilities are zero, whereas inside the bounds, the priors are not overly restrictive. 

First, we successively add observational datasets to the likelihood: 1) the $\rm{GSMF}$ alone to 2) the $\rm{GSMF}$ and $f_\mathrm{gas}$ combined, and 3) including all three observables-- $\rm{GSMF}$, $f_\mathrm{gas}$, and $\mathrm{CGD}$. The joint likelihood $\mathcal{L}_{\text{joint}}(\theta)$, at parameters $\theta$ for each of these cases, is shown in \autoref{eq:lnlike}. 

\begin{equation}
\label{eq:lnlike}
\ln \mathcal{L}_{\text{joint}}(\theta) =
\begin{cases}
\ln \mathcal{L}_{\mathrm{GSMF}}(\theta) \\
\ln \mathcal{L}_{\mathrm{GSMF}}(\theta)
 + \ln \mathcal{L}_{f_\mathrm{gas}}(\theta) \\
\ln \mathcal{L}_{\mathrm{GSMF}}(\theta)
 + \ln \mathcal{L}_{f_\mathrm{gas}}(\theta)
 + \ln \mathcal{L}_{\mathrm{CGD}}(\theta)
\end{cases}
\end{equation}

Each component $\mathcal{L}_{\mathrm{GSMF}}(\theta)$, $\mathcal{L}_{\rm{f}_{\rm gas}}(\theta)$ and $\mathcal{L}_{\mathrm{CGD}}(\theta)$ here is a Gaussian Likelihood that quantifies the agreement between the emulator-predicted observables and the observed data, weighted by data uncertainties. The posterior probability combines the likelihood and prior probability, and the surface is explored via MCMC. We use the affine-invariant MCMC Ensemble sampler \citep{Goodman2010} implemented in Python by \cite{emcee} to draw the samples from the posterior surface. For each MCMC run, 
100 walkers are initialized within the prior bounds using uniform sampling or Gaussian perturbations around a mean value. We first run 100 chain burn-in runs and then 1000 evaluations sampled from the posterior distribution. With the resulting samples, we can also find the \textit{best-fit} parameters by calculating the median of the samples.

\begin{figure*}
    \includegraphics[width=1\textwidth]{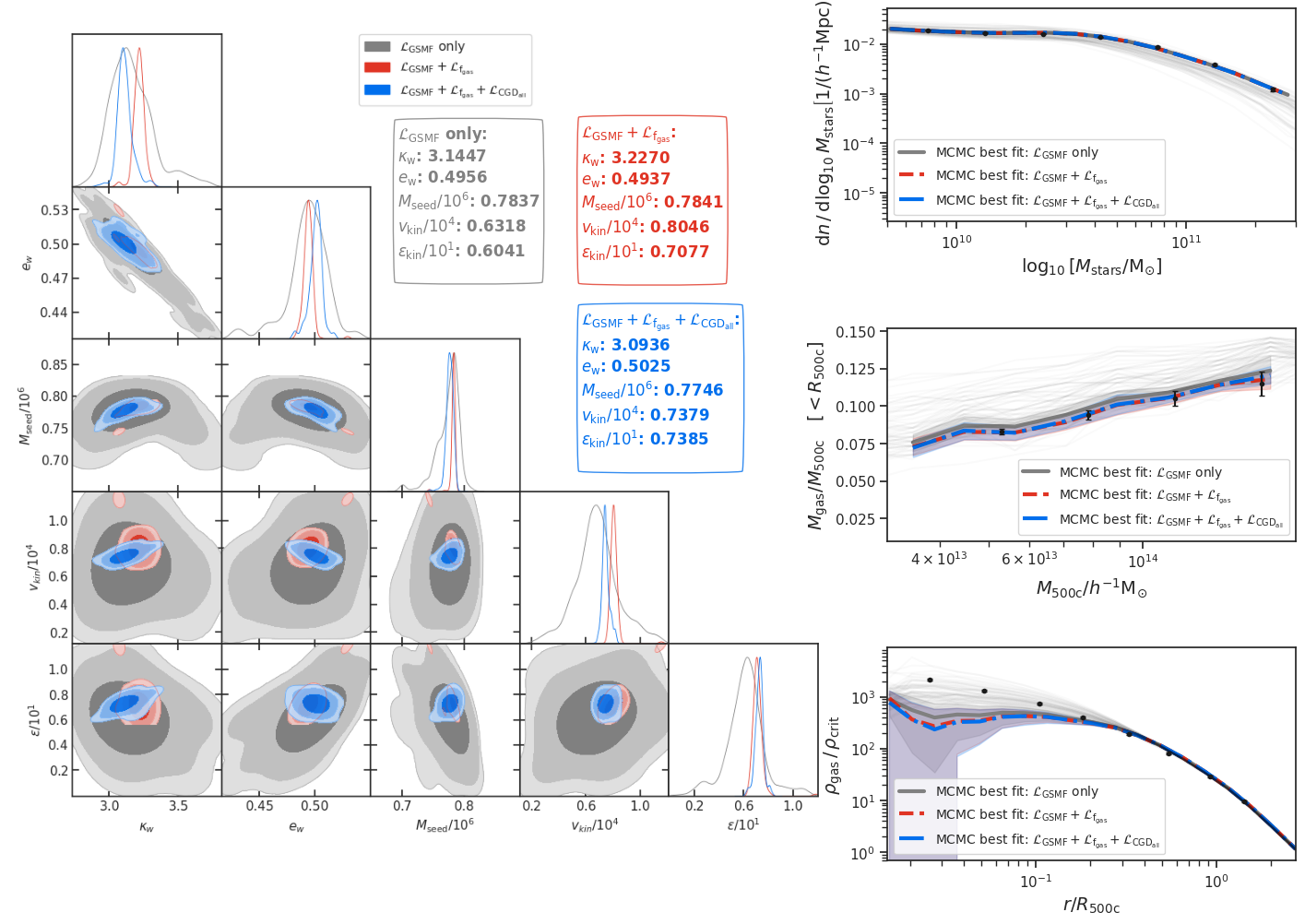}
    \caption{Right panel: Bayesian posterior distribution for the five subgrid parameters obtained via MCMC sampling in Phase-1. The forward model involves three combinations of the $\rm{GSMF}$, $f_\mathrm{gas}$, and $\mathrm{CGD}$. The plots show the effect using $f_\mathrm{gas}$ targets as a part of the calibration, which results in high feedback and deteriorated inner profiles of the $\mathrm{CGD}$ in the clusters. Left panels show the emulator predictions (solid gray, red, and blue lines) along with the errors corresponding to 5th and 95th percentiles. Black dots show the observational datasets listed in \autoref{sec:obs}. Faint gray lines show the ensemble of training summary statistics.}
    \label{fig:mcmc3b}
\end{figure*}

\begin{figure*}
    \includegraphics[width=1\textwidth]{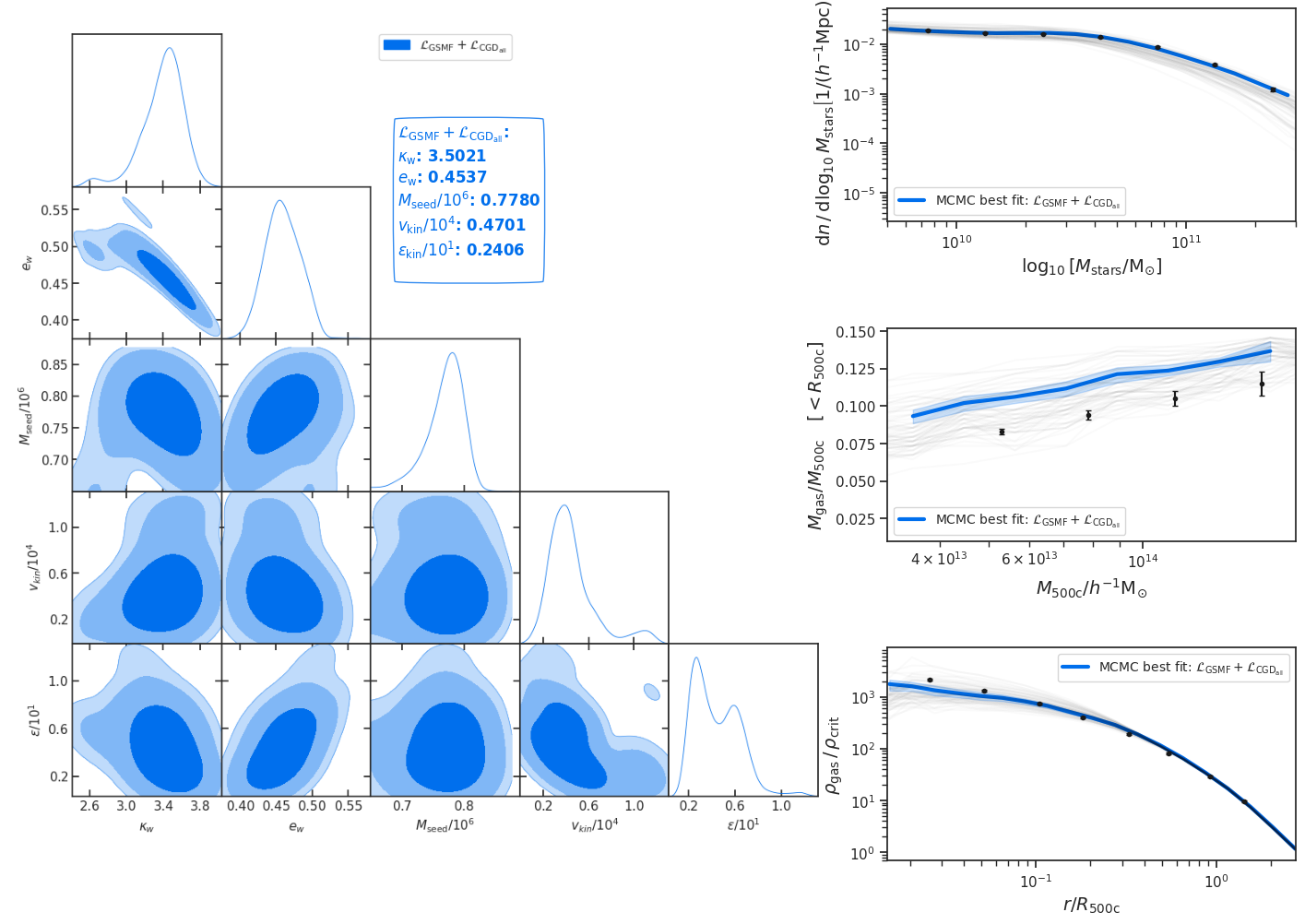}
    \caption{The MCMC posterior distribution with likelihood function that includes $\rm{GSMF}$ and $\mathrm{CGD}$, but not $f_\mathrm{gas}$. The Phase-1 emulators are used in the likelihood calculation. Right panel: Comparison of the emulator predictions at the best-fit parameters (solid blue lines) against the observational data compiled in \autoref{sec:obs} (black dots). Faint gray lines show the ensemble of training summary statistics. The emulation results show an agreement with observed $\mathrm{CGD}$ profiles, but the $f_\mathrm{gas}$ predictions are higher than the observational data.}
    \label{fig:mcmc_cgd}
\end{figure*}

We examine how joint-likelihood calibration across multiple observables improves subgrid-physics constraints. When a single observable and a corresponding target observation are used in the likelihood, the degeneracies in parameter space may not be resolved. This is due to the parameters corresponding to different subgrid physics and their manifestation in the summary statistics in a complex manner. For instance, the star formation rates are regulated by the galactic outflows, which affect the $\rm{GSMF}$. On the other hand, including $\mathrm{CGD}$ profiles provides tighter constraints for the AGN model. 
\autoref{fig:mcmc3b} demonstrated the advantages of carrying out a joint likelihood analysis across multiple observables in subgrid physics calibration. In each of the panels, the posteriors obtained from $\rm{GSMF}$ alone (gray shaded regions) are the least constraining across all parameters. This is particularly evident in the case of the kinetic parameters $\epsilon_\text{kin}$ and $v_\text{kin}$, which are known to have the least effect on the galaxy stellar masses. With the addition of $f_\mathrm{gas}$ and $\mathrm{CGD}$ (red and blue contours in \autoref{fig:mcmc3b} respectively), the posterior distributions are tighter, signifying the effect of adding cluster-level statistics in the calibration. 

Throughout this approach of successively enhancing the likelihood, several parameters remain consistently constrained: $\kappa_\text{w}$ is roughly constrained around 3, $e_\text{w}$ to approximately 0.5, and the seed mass $M_\text{seed}$ converges to $0.8 \times 10^6~M_\odot h^{-1}$ (Best-fit parameters shown in the text boxes of \autoref{fig:mcmc3b}.) The AGN kinetic feedback parameters exhibit a distinctive pattern across all likelihood combinations. The kinetic feedback efficiency $\epsilon_\text{kin}$ consistently points toward high values ranging between 6 and 7.4, while the feedback velocity $v_\text{kin}$ varies between $0.6 \times 10^4~\rm{km/s}$ and $0.8 \times 10^4~\rm{km/s}$.

When cluster-level statistics are excluded—i.e., when the $\rm{GSMF}$ is the sole component in the likelihood—the best-fit parameters for the galactic wind model and black hole seed mass are given in \autoref{tab:params3fix}. As shown in the sensitivity analysis (\autoref{fig:sensitivity}, left panel), the $\rm{GSMF}$ is primarily sensitive to these three parameters and exhibits minimal variation with the kinetic feedback parameters $v_\text{kin}$ and $\epsilon_\text{kin}$. We adopt these $\rm{GSMF}$-only values as the baseline for Phase-1, since the cluster-level observables ($f_\mathrm{gas}$ and $\mathrm{CGD}$) principally constrain the kinetic feedback sector rather than the wind parameters.

\begin{table}[ht]
\centering
\begin{tabular}{lcc}
\toprule
\textbf{$\theta_{\text{sub}}$} & \textbf{Best-fit values} \\ 
\midrule
$\kappa_\text{w}$ & $3$  \\
$e_\text{w}$ & $0.5$  \\
$M_\text{seed} [M_\odot h^{-1}]$ & $0.8 \times 10^6$  \\ \bottomrule
\end{tabular}
\caption{Best-fit values for the wind parameters and black hole seed mass from Phase-1, based on the $\rm{GSMF}$-only likelihood. Values are rounded to significant figures.}
\label{tab:params3fix}
\end{table}

In the three right panels of \autoref{fig:mcmc3b}, we also show the emulated observables at the MCMC best-fit subgrid parameters, along with the observational targets. These best-fit predictions reveal a tension: While the emulated $\rm{GSMF}$ and $f_\mathrm{gas}$ at the best fit point overlap with the target observations in the top and the middle right panels, the $\mathrm{CGD}$ shows a clear departure from the observations at $r/R_{\rm 500c} < 0.2$. Interestingly, this behavior is seen even when the $\mathrm{CGD}$ is a part of the likelihood, indicating a lower relative importance of the $\mathrm{CGD}$ in the likelihood compared to other metrics. This systematic mismatch suggests potential limitations in our ability to simultaneously reproduce all observational constraints with the current subgrid models. 

To find a set of subgrid physics parameters that better match the $\mathrm{CGD}$ profiles at the inner region of massive clusters, we conduct an additional set of MCMC runs using only $\ln \mathcal{L}_{\mathrm{GSMF}}(\theta) + \ln \mathcal{L}_{\mathrm{CGD}}(\theta)$, deliberately excluding $f_\mathrm{gas}$ from the optimization process. This modified approach yields significantly different kinetic feedback parameters while maintaining consistency in the wind parameters, as shown in \autoref{fig:mcmc_cgd}. The feedback efficiency decreases substantially from approximately $7$ to $2.4$, accompanied by a reduction in the feedback velocity to $v_\text{kin}$ approximately $0.5 \times 10^4~\rm{km/s}$, while $\kappa_\text{w}$, $e_\text{w}$, and $M_\text{seed}$ remain roughly unchanged. Under this alternative calibration scheme, the $\rm{GSMF}$ achieves a similar agreement with observations as \autoref{fig:mcmc3b}, whereas the $\mathrm{CGD}$ has a much closer agreement with the target observations. However, the emulated gas fraction at these best-fit parameters, which is not used in this particular calibration, is systematically overestimated compared to the observed gas fractions. This trade-off highlights the tension between different observational constraints in our modeling framework. This trade-off is seen with sensitivity plots in \autoref{fig:sensitivity} as well: a way to obtain higher values of the $\mathrm{CGD}$ values at inner profiles (without affecting the $\rm{GSMF}$) is to reduce $\epsilon_\text{kin}$. However, this simultaneously increases the $f_\mathrm{gas}$ values. 

To summarize, this phase of the calibration reveals two competing solution models: When constrained using $f_\mathrm{gas}$, our emulator-based inference favors stronger kinetic feedback, consistent with the observed baryon fractions but producing underdense cores for massive clusters. In contrast, a calibration that is driven by $\mathrm{CGD}$ shifts the preference toward weaker feedback, improving agreement with the observed radial gas distributions. 
A caveat is that the $\mathrm{CGD}$ measurement is sample-limited in the $(128\,h^{-1}\text{Mpc})^3$ simulation volume due to the small number of massive clusters, motivating the larger-volume study in the next section. In contrast, the $f_\mathrm{gas}$ measurement is less prone to small-number statistics, since the relevant mass range is well sampled in the current volume.

\subsection{Phase-2: Updated Simulation Design, Emulators and Calibrations}
\label{sec:phase2}

Limited simulation volumes introduce biased or noisy measurements of a few astrophysical observables, particularly when the statistics involve massive clusters. On average, the simulations from Phase-1 (with a limited volume of $V = (128\,h^{-1}\text{Mpc})^3$) effectively have no clusters above a mass of $10^{14.5}$\,M$_\odot$, and fewer than ten clusters above a mass of $10^{14}$\,M$_\odot$. This limitation impacts profile-based summary statistics like the $\mathrm{CGD}$. While larger simulation volumes improve this measurement, the computational cost of a simulation design across five subgrid parameters poses a significant constraint. To quantify this volume-dependent bias, we estimated a $\mathrm{CGD}$ correction factor using a single larger-volume simulation and performed corresponding calibration tests. This exercise, presented in \autoref{sec:appendix_disc}, further motivated the larger-volume simulation campaign described below.

Noting that 1) the primary motivation for a larger simulation volume is to get better $\mathrm{CGD}$ estimates and to utilize the constraining power of X-ray gas measurements, 2) the $\mathrm{CGD}$ variation is the highest with the two kinetic feedback parameters and 3) variations in the velocity of AGN feedback events $v_\text{kin}$ and the black hole accretion rate efficiency $\epsilon_\text{kin}$ have limited effects on galaxy stellar masses, we run a new suite varying only two parameters. For this suite, the galactic outflow parameters and black hole seed mass are fixed to the Phase-1 best-fit values given in \autoref{tab:params3fix}: $\kappa_\text{w} = 3$, $e_\text{w} = 0.5$, and $M_\text{seed} = 0.8 \times 10^6~M_\odot h^{-1}$. We restrict this exploration to the weak kinetic AGN feedback solution in Phase-1, the regime where the simulated $\mathrm{CGD}$ profiles are in better agreement with observational targets. Thus, the range of the two kinetic feedback parameters will also be restricted based on the posterior distribution estimates in the Phase-1 studies, as shown in \autoref{tab:params2}. 

\begin{table}[ht]
\centering
\begin{tabular}{lcc}
\toprule
\textbf{$\theta_{\text{sub}}$} & \textbf{min($\theta_{\text{sub}}$)} & \textbf{max($\theta_{\text{sub}}$)} \\ 
\midrule
$v_\text{kin} [\rm{km/s}]$ & $0.3 \times 10^4$ & $1.0 \times 10^4$ \\
$\epsilon_\text{kin}$ & $0.2$ & $3$ \\ \bottomrule
\end{tabular}
\caption{Parameter bounds for kinetic feedback parameters in the experimental design of Phase-2, showing the minimum and maximum values for each parameter}
\label{tab:params2}
\end{table}

The above considerations allow us to limit the new simulation suite to just 16 models with a larger side-length of $L = 256\,h^{-1}\text{Mpc}$. The mass resolution of the simulations in Phase-2 remains the same as in Phase-1. Due to the large boxes, we now have increased cluster counts in the simulations (roughly 10 massive clusters with mass $10^{14.5}$\,\,M$_\odot$). From each simulation, we identify the clusters (with SOD Mass $M_{500c} > 3 \times 10^{14}\, h^{-1}M_\odot$) and compute the $\mathrm{CGD}$. The Phase-2 emulator is then created to map the two kinetic feedback modeling parameters to the $\mathrm{CGD}$. The emulator construction and validation are carried out in the same way as in Phase-1. With the increased cluster statistics, the Phase-2 emulator achieves a mean error of approximately 3\% for the $\mathrm{CGD}$—a substantial improvement over the 9\% error in Phase-1.

\subsubsection{Constraints from Phase-2 Simulations}

\begin{figure*}
\includegraphics[width=1.0\textwidth]{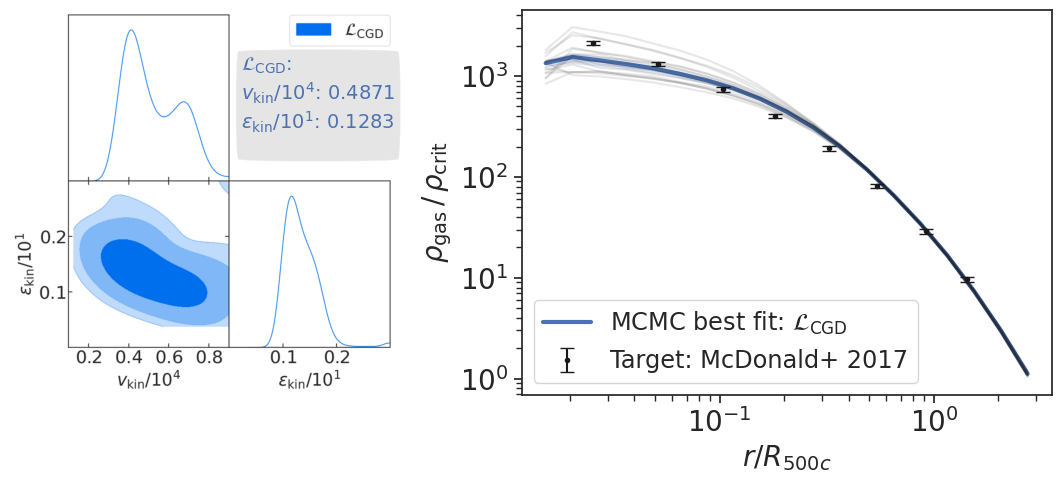}
\caption{Phase-2 constraints on the kinetic feedback parameters. Left panel: Bayesian posterior distribution of two subgrid parameters obtained using MCMC sampling. The blue contour shows the posterior corresponding to just the  $\mathrm{CGD}$ (from the larger box). The text box lists the best-fit kinetic feedback parameters. Right panels: Emulated $\mathrm{CGD}$ profiles (blue lines) corresponding to the best-fit parameters with the target observational datapoints (black dots).}
\label{fig:mcmc2p}
\end{figure*}

With the newly trained emulator for the $\mathrm{CGD}$, we run a second Bayesian inference against target observations. Both the number of parameters and the ranges are restricted compared to Phase-1. The target astrophysical measurement in Phase-2 is limited to $\mathrm{CGD}$ profiles alone, since the dependence of the $\rm{GSMF}$ and $f_\mathrm{gas}$ on just the kinetic feedback parameters is limited when the galactic wind model is fixed. 

The Bayesian posterior distribution further restricts the kinetic feedback parameters. The best-fit kinetic feedback velocity is $v_\text{kin} \approx 0.5 \times 10^4~\rm{km/s}$, and the AGN feedback efficiency decreases further (compared to Phase-1) to $1.3$, as shown in the left panel of \autoref{fig:mcmc2p}. The right panel in the plot shows the emulator prediction of the $\mathrm{CGD}$ profiles in agreement with the target observations from \cite{mcdonald2017remarkable}.

With this two-phase posterior estimation process, we obtained two sets of parameters that are calibrated across multiple observations. One set of the best-fit parameters corresponds to a high velocity of the feedback events and a high kinetic feedback efficiency. However, to get better agreement with $\mathrm{CGD}$ profiles, we explore another joint calibration configuration that excludes the $f_\mathrm{gas}$, resulting in low kinetic AGN feedback and a better fit to the cluster profiles. This modality is explored further in Phase-2 with a larger simulation that consists of a significantly larger number of massive clusters. This mode demonstrates a further reduction of kinetic feedback efficiency $\epsilon_{\rm{kin}}$. 

The best-fit kinetic feedback parameters for both calibration regimes are summarized in \autoref{tab:params2fix}. The high-feedback configuration is derived from the Phase-1 posterior using the joint likelihood $\mathcal{L}_{\mathrm{GSMF}}(\theta) + \mathcal{L}_{f_\mathrm{gas}}(\theta)$, which favors stronger AGN feedback to match observed cluster gas fractions. The low-feedback configuration is obtained from Phase-2, where the parameters in \autoref{tab:params3fix} are held fixed and only $\mathcal{L}_{\mathrm{CGD}}(\theta)$ is used, favoring weaker feedback to reproduce the observed inner gas density profiles.

\begin{table}[ht]
\centering
\begin{tabular}{lcc}
\toprule
\textbf{$\theta_{\text{sub}}$} & 
\parbox{2.5cm}{\centering \textbf{Best-fit values} \\ \textbf{(high feedback)}} & 
\parbox{2.5cm}{\centering \textbf{Best-fit values} \\ \textbf{(low feedback)}}\\
\midrule
$v_\text{kin}[\rm{km/s}]$ & $0.8 \times 10^4$ & $0.5 \times 10^4$ \\
$\epsilon_\text{kin}$ & $7.1$ & $1.3$ \\ \bottomrule
\end{tabular}
\caption{Best-fit kinetic feedback parameters for the two calibration regimes. The high-feedback configuration, calibrated against $f_\mathrm{gas}$, produces lower gas fractions but underpredicts inner $\mathrm{CGD}$ profiles. The low-feedback configuration, calibrated against $\mathrm{CGD}$, better reproduces cluster core densities but overpredicts gas fractions compared to the corresponding observational dataset.}
\label{tab:params2fix}
\end{table}

We note that potential bias terms -- such as those associated with cosmic variance and with uncertainties in cluster mass calibration -- have not been treated explicitly in the discussion above. A recent analysis by \cite{kugel2023flamingo} incorporated three bias parameters directly in the calibration and found that biases affecting the $\rm{GSMF}$, including systematic effects from finite survey volumes and stellar-mass inference, are negligible. In contrast, they report a statistically significant bias associated with the underestimation of cluster masses when masses are derived under the assumption of hydrostatic equilibrium. We repeat an analogous exercise in our framework, adopting a bias prescription closely aligned with \cite{kugel2023flamingo}; details and results are presented in \autoref{sec:appendix_bias}. Our experiments yield qualitatively similar conclusions, in agreement with the findings of \cite{kugel2023flamingo}.

\section{Additional Measurements}
\label{sec:addon}

\begin{figure*}[t]
    \centering
    \includegraphics[width=0.32\linewidth]{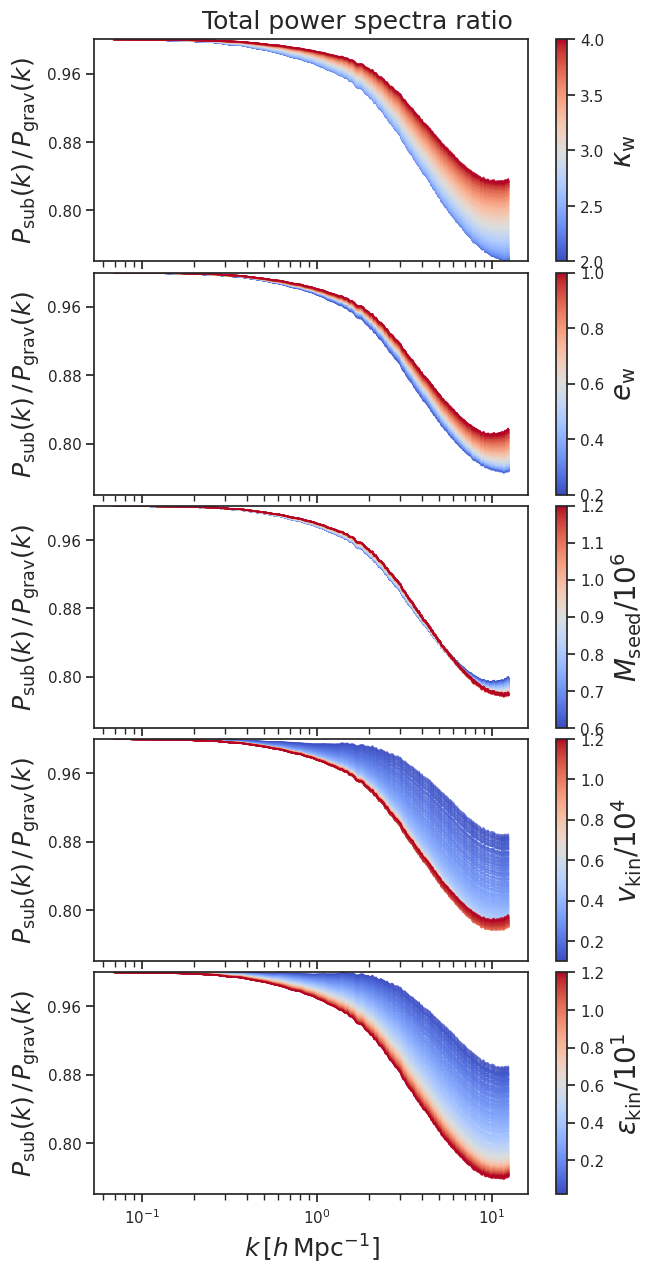}
    \includegraphics[width=0.31\linewidth]{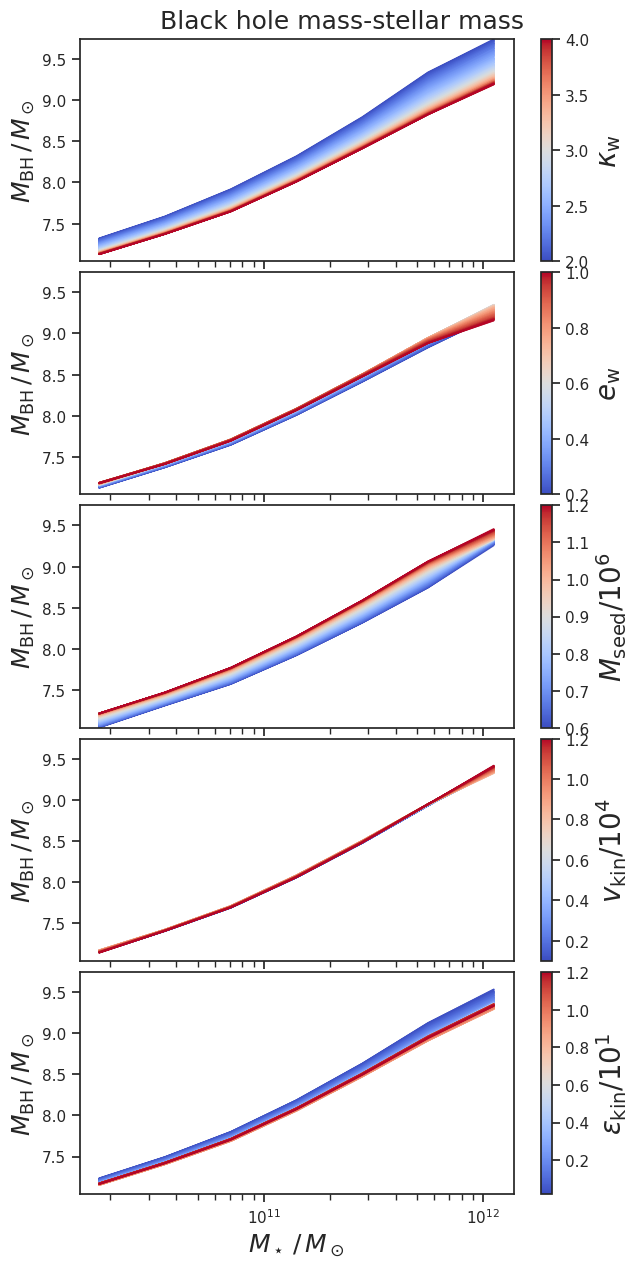}
    \includegraphics[width=0.33\linewidth]{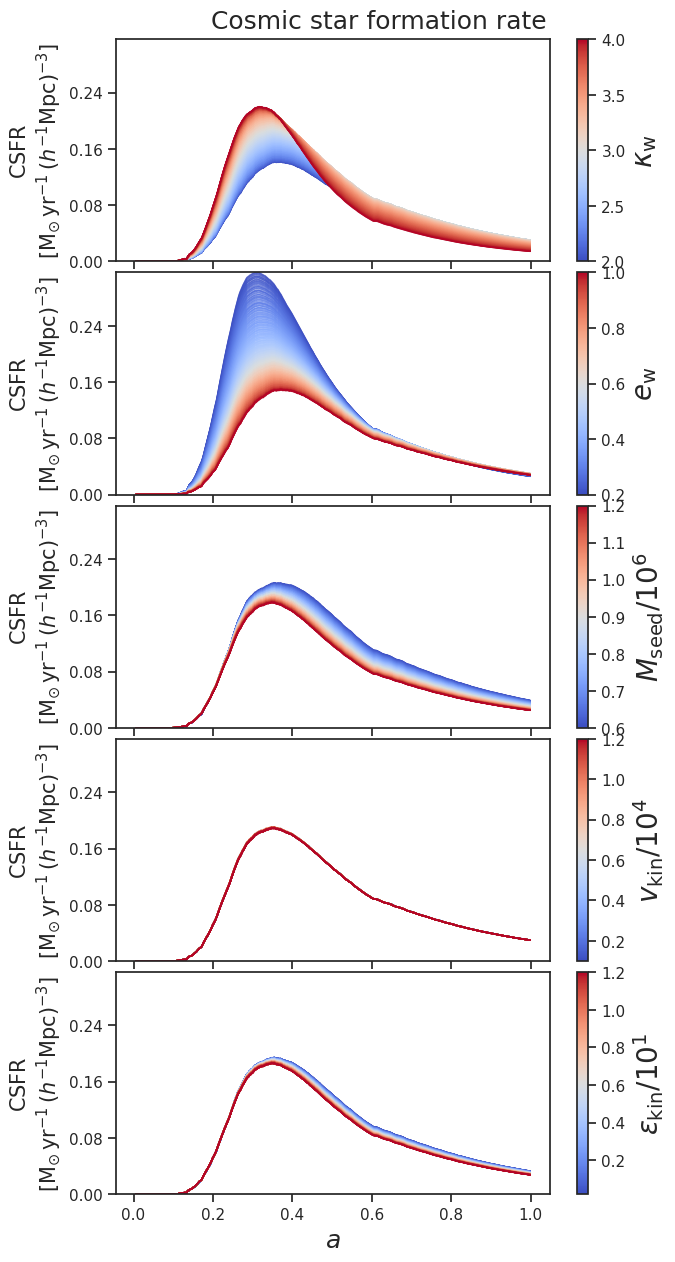}
    \caption{Results for the sensitivity analysis for the three observables not used in parameter calibration. The impact of five subgrid parameters is investigated. Within each panel, the variation of a summary statistic is shown when a single parameter is varied, while the rest are fixed to the center of the experimental design.}
    \label{fig:sensitivity_pk}
\end{figure*}

In addition to the observables used for the parameter calibration in \autoref{sec:calibration}, we emulate three additional summary statistics. First, we emulate the ratio $P_{\rm sub}(k)/P_{\rm grav}(k)$ of the total matter power spectra from hydrodynamic simulations to their GO counterparts with identical initial conditions. The $k$-range for these simulations is bounded by the fundamental mode set by the box size, $k_{\rm min}=2\pi/L$, and by the particle Nyquist wavenumber, $k_{\rm max}=\pi/(L/N_{\rm p}^{1/3})$, where $N_{\rm p}$ is the total number of dark matter (or baryonic tracer) particles and $L/N_{\rm p}^{1/3}$ is the mean inter-particle spacing in the initial conditions. For our Phase-1 simulations, this yields $k_{\rm min}\simeq 0.05\,h\,\mathrm{Mpc}^{-1}$ and $k_{\rm max}\simeq 12.57\,h\,\mathrm{Mpc}^{-1}$.

We also create an emulator for the black hole mass to stellar mass relationship, $\rm{BHMSM}$. This relationship quantifies the correlation between the mass of the supermassive black hole at the center of a galaxy and the stellar mass of the galaxy. It reflects the co-evolution of galaxies and their central black holes, driven by processes like AGN feedback, galaxy growth, and mergers \citep{kormendy2019coevolution, mcconnell2013revisiting}. In our simulation outputs, the $\rm{BHMSM}$ covers the stellar mass range $10^{10} \, M_\odot \leq M_\star\,/\,M_\odot \leq 2 \times 10^{12} \, M_\odot$.

In addition, we build an emulator for the cosmic star formation rate ($\rm{CSFR}$), which traces the evolution of the star formation rate density across cosmic time. The $\rm{CSFR}$ provides critical insights into how stellar mass is changed in galaxies, shaped by the interplay of gas accretion, feedback mechanisms, and large-scale structure formation (see a recent review by \citealt{madau2014cosmic}). The emulator errors for these additional observables are less than 1\% for the $\rm{BHMSM}$, approximately 4\% for the $\rm{CSFR}$, and 1.6\% for $P_{\rm sub}(k)/P_{\rm grav}(k)$ in Phase-1. 

While these emulators are not used for our calibration process, they provide insight into the effect of subgrid models. In \autoref{fig:sensitivity_pk}, the power spectra suppression is close to unity in large scales ($k < 0.2 h$Mpc$^{-1}$, indicating a minimal effect of the baryonic processes on the matter distribution at large scales. On smaller scales ($k > 0.2 h$Mpc$^{-1}$, the feedback effects are more pronounced, dropping the dark matter power spectra $P_{\rm sub}(k)$ to around 80\% of that of the gravity-only counterpart. Notably, increasing the wind velocity parameter and the wind energy injection parameter results in less power suppression, i.e., the ratio approaches unity. In contrast, higher values of the AGN feedback parameters $v_{\rm kin}$ and $\epsilon_{\rm kin}$ lead to stronger power suppression. Additionally, the impact of the black hole seed mass is relatively minor, with little variation in power suppression observed as $M_\text{seed}$ is adjusted. 

On the other hand, $\rm{BHMSM}$ responds most strongly to the wind velocity and black-hole accretion efficiency, and only weakly to $v_{\rm kin}$ and $e_w$, as shown in \autoref{fig:sensitivity}. Decreasing wind velocity and the black-hole accretion rate efficiency yield more massive black holes, whereas increasing the seed mass $M_\text{seed}$ shifts the relation in the opposite direction. In contrast to $\rm{BHMSM}$, the $\rm{CSFR}$ displays the highest variation with the wind parameters $\kappa_w$ and $e_w$, and the least variation with the kinetic feedback parameters. 

\subsection{Simulation at Optimal Parameters}

After inferring plausible subgrid model parameters using our two-phase calibration approach, we conduct an additional simulation to further validate the calibration method. We use the same mass resolution as our simulation suites, with side-length $L = 256\,h^{-1}$Mpc. The subgrid parameters are fixed at $\kappa_\text{w}=3$, $e_\text{w}=0.5$, $M_\text{seed}/10^6 = 0.8$, $v_\text{kin}/10^4=0.51$, $\epsilon_\text{kin}/10^1=0.13$, where $\kappa_\text{w}$, $e_\text{w}$, and $M_\text{seed}$ are set to their Phase-1 calibrated values (\autoref{tab:params3fix}) and $v_\text{kin}$ and $\epsilon_\text{kin}$ follow the low-feedback configuration refined in Phase-2 (\autoref{tab:params2fix}). 
This simulation, Frontier-E-Small, is a precursor to a survey-scale hydrodynamic simulation \cite{frontiere2025cosmological}, carried out to validate and quantify the choices of subgrid models, the agreement of summary statistics with observational targets, as well as other criteria. It was also used for comparison with modern hydrodynamic simulations in \cite{frontiere2025modeling}.

\begin{figure*}[t]
    \centering
    \includegraphics[width=0.99\linewidth]{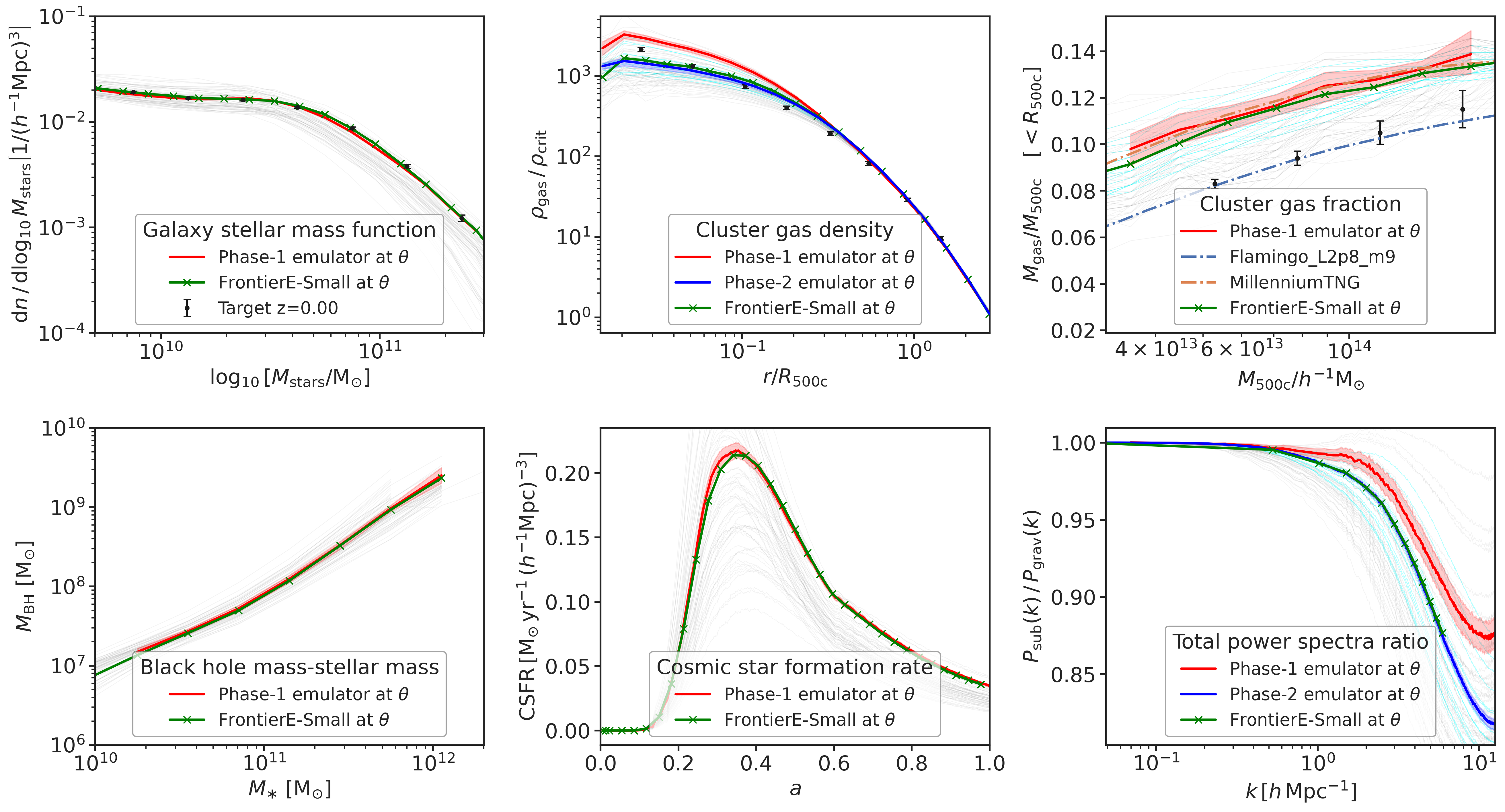}
    \caption{Emulators measurements compared to the Frontier-E-Small test simulation (green lines). Phase-1 emulator estimation (red lines) is shown for all the observables, whereas Phase-2 emulators are only constructed for $\mathrm{CGD}$ and $P_{\rm sub}(k)/P_{\rm grav}(k)$ ratio (shown in blue). Observational target data (black dots in the top row panels) are shown for observables used in calibration. For $f_\mathrm{gas}$ (top right panel), we also show simulation results from other simulation efforts. The faint gray lines correspond to the training summary statistics from Phase-1, and the faint blue lines correspond to the summary statistics from Phase-2.}
    \label{fig:fEsmall}
\end{figure*}

The results from the Phase-1 and Phase-2 emulators, along with associated error bars (corresponding to 5th and 95th percentiles), are shown in \autoref{fig:fEsmall}. For each of the summary statistics, we also show the results from the Frontier-E-Small simulation. Whenever applicable, observational targets and other comparable simulation results are also shown. 

The top left panel in \autoref{fig:fEsmall} shows the $\rm{GSMF}$ predicted by the Phase-1 emulator closely matching both the observational data and the Frontier-E-Small simulation for $z=0$. In the top middle panel, the $\mathrm{CGD}$ prediction from Phase-2 matches closely with observational targets and the Frontier-E-Small simulation measurements. The difference between Phase-1 and Phase-2 measurements of the $\mathrm{CGD}$ is primarily due to cluster number counts, as explained in \autoref{sec:phase2}. 

Comparison of $f_\mathrm{gas}$ measurements from multiple sources is shown in the top right panel of \autoref{fig:fEsmall}. First, we note that the observational dataset used for the ML-based subgrid parameter calibration in \citet{kugel2023flamingo} is consistent with the $f_\mathrm{gas}$ measurements reported for the cosmology-scale FLAMINGO runs in \citet{schaye2023flamingo}, which adopt the calibrated parameter set.
In this regime, large fractions of gas are expelled from cluster cores, yielding $f_\mathrm{gas}\approx0.08$ at $M_{500c}\sim 6\times10^{13} M_\odot/h$. This results in low $f_\mathrm{gas}$ values, but higher cluster gas densities at the inner radii (as seen in \autoref{sec:phase1}). On the other hand, the low–kinetic-feedback mode, selected by matching the $\rm CGD$ profiles, implies a different regime: the resulting $f_\mathrm{gas}$ predictions are inconsistent with the gas-fraction compilation of \citet{kugel2023flamingo}, and instead indicate substantially higher gas retention, with $f_\mathrm{gas}\gtrsim 0.12$ for $M_{500c}\gtrsim 10^{14},M_\odot/h$. In other words, kinetic jets in the FLAMINGO simulation are more effective at suppressing gas buildup in cluster centers than our feedback models. We also note that the Millennium-TNG and Frontier-E-Small simulations are within our error estimates of our emulator predictions of $f_\mathrm{gas}$. 

The Phase-1 emulators for the $\rm{BHMSM}$ and $\rm{CSFR}$ also agree closely with the Frontier-E-Small simulation (bottom left and bottom middle panels of \autoref{fig:fEsmall} respectively), indicating that—similar to the $\rm{GSMF}$—these statistics are largely insensitive to the simulation volume over the ranges considered. 

The total matter power spectrum suppression, $P_{\rm sub}(k)/P_{\rm grav}(k)$, is another observable for which finite-volume effects can be significant. As shown in the bottom-right panel of \autoref{fig:fEsmall}, the Frontier-E-small simulation exhibits stronger suppression compared to the Phase-1 emulator prediction, where the emulator is trained on a volume that is $1/8$-th of the simulation volume. In simulations with side-length $128\,h^{-1}{\rm Mpc}$, the global $P_{\rm sub}(k)/P_{\rm grav}(k)$ ratio can be dominated by a small number of the massive clusters. This occurs because the kinetic AGN feedback that drives the suppression is activated primarily in these high-mass environments. By contrast, lower-mass clusters experience little or none of this kinetic mode and therefore contribute only weakly to the aggregate signal. The sensitivity of $P_{\rm grav}(k)$ to the choice of random seed is well documented (e.g., Section~3.3 of \citealt{heitmann2013coyote}), and similar seed-driven variance has more recently been demonstrated for the baryonic suppression (e.g., Appendix~A of \citealt{bigwood2025case}). Consistent with this interpretation, re-training the emulator on the larger-volume Phase-2 suite (side-length $256\,h^{-1}{\rm Mpc}$) yields predictions (with emulator error of less than 1\%) that closely match the reference Frontier-E-small simulation.

For both the $\mathrm{CGD}$ and $P_{\rm sub}(k)/P_{\rm grav}(k)$, we verified that the volume dependence observed between Phase-1 and Phase-2 reflects insufficient cluster sampling rather than a fundamental sensitivity to box size. Our comparisons with the survey-scale Frontier-E simulation \citep{frontiere2025cosmological}, which employs a $(4.655\,\rm{Gpc})^3$ volume, confirm that results remain stable once sufficient cluster statistics are achieved.

\section{Discussions}
\label{sec:conclusions}

Calibrating subgrid model parameters in cosmological hydrodynamic simulations is critical for obtaining accurate observational predictions. In this work, we leveraged the CRK-HACC simulation framework, combined with emulator-based inference, to obtain the best-fit parameters (Tables~\ref{tab:params3fix}~and~\ref{tab:params2fix}) for galactic outflows and AGN models that correspond to multi-probe observations. The tuned subgrid models serve two key scientific purposes: 1) They provide insights into how astrophysical processes and their effects manifest on galactic scales,  and 2) they provide a robust framework for determining best-fit subgrid parameters for survey-scale hydrodynamic simulations. 

In CRK-HACC, gas stochastically converts into wind and star particles at rates governed by scaled wind velocity and energy injection parameters. In addition, AGN are seeded within galaxies, with thermal (dictated by the seed mass of the black holes) and kinetic feedback processes (governed by velocity and accretion rate efficiency). One can constrain all these subgrid parameters from observational data. For robust tuning, we use number statistics like the $\rm{GSMF}$, cluster-level properties of the amount of gas, and their radial profiles for different core populations. In this work, we restrict our analysis to a single mass resolution. Extending the calibration framework to multiple resolutions would require accounting for resolution-dependent stochasticity and local environmental effects on cluster properties.

Fast surrogate models for simulation products involving clustering statistics like power spectra have been successful in achieving sub-percent accuracy with meticulous experimental designs, GP interpolation, and Bayesian inference. We use the familiar methodology in the case of astrophysical observables, with a novel multi-stage approach of using an initial emulator suite for the calibration (in Phase-1), and a second stage of emulators for refining the predictions. We have shown that multiple simulation fidelities (in our case, two different simulation volumes) result in considerable improvements in the calibration process. The two-phase approach adopted here was driven by computational constraints and the need for improved cluster statistics. More sophisticated strategies—such as adaptive experimental designs or active learning—could further optimize the allocation of simulation resources in future calibration efforts.

In transitioning from GO to hydrodynamic simulations, the connection between simulations and observations becomes more direct—hydrodynamic simulations produce baryonic observables such as galaxy stellar masses and gas fractions without requiring intermediate modeling steps like the galaxy-halo connection. At the same time, the observables used for calibration (e.g., cluster gas fractions, radial profiles) probe smaller scales where observational uncertainties and intrinsic scatter are larger, allowing for a higher error budget than the sub-percent requirements often encountered for large-scale matter power spectrum predictions. However, due to unresolved physical processes in hydrodynamic simulations, numerical predictions are also associated with larger uncertainties, complicating model validation. Moreover, different subgrid models can produce similar galaxy populations, adding to the degeneracies in the subgrid parameter fits.

Another issue is the size of the simulation volumes. In our comparisons of multiple simulation volumes, we see that the lack of large clusters results in biases in radial profile aggregates and power spectra suppression. We demonstrate that there are considerable advantages to building hydrodynamic simulation suites of large volumes with more realistic cluster populations that enable the use of X-ray probes. Calibration campaigns that rely solely on small-box ensembles may therefore inadvertently bias AGN feedback parameters when profile-based or power spectrum statistics are used as calibration targets.

A central result of this work is the identification of two distinct AGN kinetic feedback regimes (tabulated in \autoref{tab:params2fix}), each favored by different observational constraints. The high-feedback configuration ($v_\text{kin} \approx 0.8 \times 10^4~\rm{km/s}$, $\epsilon_\text{kin} \approx 7$), calibrated against cluster gas fractions, efficiently expels gas from cluster cores but systematically underpredicts inner gas densities. The low-feedback configuration ($v_\text{kin} \approx 0.5 \times 10^4~\rm{km/s}$, $\epsilon_\text{kin} \approx 1.3$) reproduces observed $\mathrm{CGD}$ profiles of massive clusters, but overpredicts integrated gas fractions. This tension is not an artifact of emulator accuracy or simulation volume—it reflects a limitation in the ability of current subgrid prescriptions to simultaneously match both integrated and spatially resolved cluster gas observables. Resolving this discrepancy may require modifications to the functional form of the kinetic feedback model, additional physical processes, or a reassessment of systematic biases in the observational datasets themselves.

We conclude by outlining several avenues for further investigation:

\begin{enumerate}
    \item The current analysis is performed at a single mass resolution. This implies that the calibrated parameters are only applicable to simulations with consistent mass resolutions.  This limitation also highlights the potential for future studies to explore multi-fidelity, multi-objective optimization frameworks, which could provide deeper insights into the behavior of subgrid models across varying mass resolutions.


    \item The present calibration focuses exclusively on redshift $z=0$ datasets, omitting any redshift evolution of the observables or subgrid parameters. In future work, we will extend this analysis to include redshift-dependent constraints, enabling the calibration of subgrid models across a broader range of cosmic epochs. In \cite{frontiere2025modeling}, we show that the $\rm{GSMF}$ is already in agreement with data at higher redshifts ($0<z<2$), which is partly attributable to the adopted subgrid model implementation and the parameter choices selected in that work and held fixed in the present study.  

    \item A crucial next step is simultaneously calibrating subgrid and cosmological parameters. This approach will require a dedicated simulation suite that incorporates both baryonic and cosmological parameter variations, allowing for explorations of fundamental physics parameters, such as dark energy and neutrino masses, in conjunction with astrophysical processes. Such a framework would provide a more holistic understanding of the interplay between small-scale baryonic physics and large-scale cosmology.
\end{enumerate}

\section*{acknowledgements}

We acknowledge the efforts of Claude-André Faucher-Giguère and Patricia Larsen for their contributions to the HACC simulation framework and associated data products. We thank Dave Higdon for insightful discussions on statistical modeling and Earl Lawrence for helpful conversations during the early stages of emulator development. Work at Argonne National Laboratory was supported under the U.S. Department of Energy contract DE-AC02-06CH11357. U.S. Department of Energy, Office of Science, Office of Advanced Scientific Computing Research and Office of High Energy Physics, Scientific Discovery through Advanced Computing (SciDAC) program. The SciDAC program also supported the work at Los Alamos National Laboratory.

The emulator uses the following Python packages: Sepia \citep{james_gattiker_2020_4048801} and Scikit-learn \citep{scikit}. The analyses performed in this paper utilize the following: Numpy and Scipy \citep{numpyscipy}, Matplotlib \citep{matplotlib}, emcee \citep{emcee} and GetDist \cite{getdist}. The final emulator suite, trained models, and relevant summary statistics dataset are provided at \url{https://github.com/nesar/subgrid_emu}. 

This research used resources of the National Energy Research Scientific Computing Center, a DOE Office of Science User Facility supported by the Office of Science of the U.S. Department of Energy under Contract DE-AC02-05CH11231. This research also used
resources of the Argonne Leadership Computing Facility, which is a DOE Office of Science User Facility supported under Contract DE-AC02-06CH11357. Additionally, this work used resources of the Oak Ridge Leadership Computing Facility, which is a DOE Office of Science User Facility supported under Contract DEAC05-00OR22725.


\appendix
\section{Gaussian Process-based emulation and calibration technique}
\label{sec:appendix_gp}

The emulation construction primarily follows the techniques used in constructing the family of CosmicEmu emulators  \citep{Heitmann2006, Habib2007,2010ApJ...715..104H,2009ApJ...705..156H,2010ApJ...713.1322L}:
The Gaussian process interpolation over the input parameters is carried out on the principal component (PC) weights of a PC representation of the output space. Here, we reiterate the steps involved in constructing and deploying the emulator. 

For each summary statistic from the CRK-HACC simulations, we first perform any necessary pre-processing steps, such as masking the valid ranges or log-scaling to deal with the dynamical ranges. Next, we apply the Singular Value Decomposition (SVD) to the scaled and centered summary statistic $\chi(k, \theta)$, a function of $k$ and subgrid parameters $\theta$, yielding a low-dimensional representation of the outputs comprised of a set of orthogonal bases and their associated basis weights. That is, an approximation of the summary statistic $\chi{'}(k, \theta)$ can be expressed as a truncated sum of orthogonal basis vectors $\phi_m(k)$ and their corresponding weights $w_m(\theta)$:
\begin{equation} 
\chi{'}(k, \theta) = \sum_{m=1}^{n_w} \phi_m(k) w_m(\theta) + \epsilon, \end{equation}
where $n_w$ is chosen to capture over 95\% of the data variance to reduce the complexity of the problem while retaining the dominant features of the data. 
This truncation drastically reduces the dimensionality from a larger $n_{bins}$ to a smaller $n_w$, making the interpolation both faster and more stable. For each of the observables, the value of $n_w$ varies (i.e., seven basis for $\rm{GSMF}$, 8 for $f_\mathrm{gas}$, 19 for $\mathrm{CGD}$ in Phase-1 (\autoref{sec:phase1}). 

A Gaussian process (GP) model will be fit to each of these sets of basis weights $w_m(\theta)$, allowing for interpolation at unobserved subgrid parameters. 
The orthogonality of the basis functions, $\phi_m(k)$, ensures that using conditionally independent GPs for each weight is appropriate. 
GPs offer a non-parametric alternative to traditional parametric methods, which estimate a finite number of model parameters from the data. Instead of fitting a predetermined functional form, GPs estimate a distribution over possible functions $f(x)$ that are consistent with the observed data. In a GP model with training inputs ${x_1, \dots, x_n}$ and corresponding targets ${y_1, \dots, y_n}$, the joint distribution of the function values is assumed to be Gaussian:
\begin{equation} p(f(x_1), \dots, f(x_n)) \sim \mathcal{N}(\mu, \mathbf{K}), \end{equation}
where $\mathbf{K}$ is a covariance matrix with elements $\mathbf{K}_{ij} = k(x_i, x_j)$, defined by a kernel function $k$ that describes the similarity between outputs as a function of their inputs (often, the distance between inputs). In our emulator, the inputs correspond to the subgrid parameters $\theta$, while the outputs are the weights $w_m(\theta)$ obtained from the PC decomposition.

The GP posterior distribution for a new test point $\theta_*$ can be derived from the joint distribution of the training and test points:
\begin{equation}\label{eq:joint} 
p(y, y_*) = \mathcal{N}\left( \mathbf{0}, \begin{bmatrix} \mathbf{K} & \mathbf{K}_{*}^{T}\\ \mathbf{K}_{*} & \mathbf{K}_{**} \end{bmatrix}
\right).
\end{equation}
where $\mathbf{K}*$ represents the covariance between the training and test points, and $\mathbf{K}{**}$ is the covariance of the test point with itself. The mean and variance of the predicted function value at the test point are then given by:
\begin{eqnarray} \mu(y_) &=& \mathbf{K}_ \mathbf{K}^{-1} y, \\
\sigma^2(y_) &=& \mathbf{K}{**} - \mathbf{K} \mathbf{K}^{-1} \mathbf{K}_*^T. \end{eqnarray}

Assuming the structure and hyperparameters of the kernel $k$ are known, this framework allows us to make fast, accurate predictions with associated uncertainties for any new subgrid parameter set $\theta_*$, completing the emulator construction. The computational efficiency of GPs, particularly when applied to reduced-dimensional data from PCA, enables rapid interpolation of subgrid models across a wide parameter space.  

In our case, the squared exponential kernel \citep{Rasmussen2006} is chosen for each $k$, selected because we expect the PC weights containing structure to be smoothly varying.
The emulation and calibration in this paper is done with SEPIA (Simulation-Enabled Prediction, Inference, and Analysis) \citep{james_gattiker_2020_4048801}, which implements the Bayesian emulation and calibration methodology described in \citet{Higdon2008}; SEPIA learns both the kernel hyperparameters and the residual variance from PC truncation.
Specifically, the kernel for the GP model for each $w_m(\theta)$ is defined as:
\begin{equation}
k_m(\theta, \theta') = \frac{1}{\lambda_m} \exp\left( - \sum_{d=1}^{p} \rho_{md} (\theta_d - \theta_d')^2 \right) + \zeta_m \delta_{\theta, \theta'},
\end{equation}
where $\lambda_m$ is the marginal precision (inverse variance) for the $m$th principal component weight, $\rho_{md}$ controls the correlation length scale in the $d$th input dimension for the $m$th weight, and $\zeta_m$ is a small nugget term added to the diagonal (i.e., when $\theta = \theta'$) to account for numerical noise or unresolved variability. Larger values of $\rho_{md}$ imply stronger penalization of variation along dimension $d$, enforcing smoother variation in $w_m(\theta)$.
The error from PC truncation is modeled as an additional Gaussian noise term, with its variance learned during calibration to account for residual variance beyond the retained PCs.

The calibration process follows the Bayesian formulation in \citet{Higdon2008}, where the goal is to learn a posterior over the parameter values $\Theta$ corresponding to the observable by reconciling simulation outputs with observational data. The physical observation at setting $\Theta$ is assumed to be:
\begin{equation}
y = \xi(\Theta) + \delta(\Theta) + \epsilon_y,
\end{equation}
where $\xi(\Theta)$ is the emulator prediction from the GP, $\delta(\Theta)$ is a model discrepancy term capturing systematic bias between the emulator and reality, and $\epsilon_y$ is measurement error.

SEPIA supports a basis expansion of the discrepancy term:
\begin{equation}
\delta(\theta) = \sum_{j=1}^{n_\delta} d_j \, v_j(\theta),
\end{equation}
where $\{d_j\}$ are user-specified discrepancy basis functions, and $v_j(\theta)$ are GP weights with their own kernels and hyperparameters, similarly structured as those for $w_m(\theta)$. These discrepancy terms ensure that model inadequacies are not mistakenly attributed to parameter miscalibration.

Multiple observables (e.g., $\rm{GSMF}$, $f_\mathrm{gas}$, $\rm CGD_{\rm all}$) can jointly inform the same subgrid parameter set $\theta$ by constructing separate emulators and discrepancy models for each observable. Let $y^{(i)}$ denote the $i$th observable's data and $\xi^{(i)}(\theta)$ its emulator prediction. Each observable contributes a likelihood term:
\begin{equation}
y^{(i)} = \xi^{(i)}(\Theta) + \delta^{(i)}(\Theta) + \epsilon^{(i)}_y,
\end{equation}
and these likelihoods are combined into a joint posterior over $\Theta$. Since each observable is conditioned on the same underlying parameter set, information is pooled naturally, with calibration favoring parameter settings that jointly agree with all observed data. This multi-observable setup helps constrain $\theta$ more robustly, especially when observables are complementary in their sensitivities.

\section{Bias terms}
\label{sec:appendix_bias}

When comparing simulations with observations, systematic biases in observational measurements must be accounted for to ensure a fair comparison. Our approach, following that of \citep{kugel2023flamingo}, incorporates three key bias terms that address common systematic effects in galaxy and cluster measurements. These biases allow us to properly map between the observed quantities and the true physical quantities predicted by our simulations, enabling a more robust calibration procedure.

\begin{figure*}
    \includegraphics[width=1.0\textwidth]{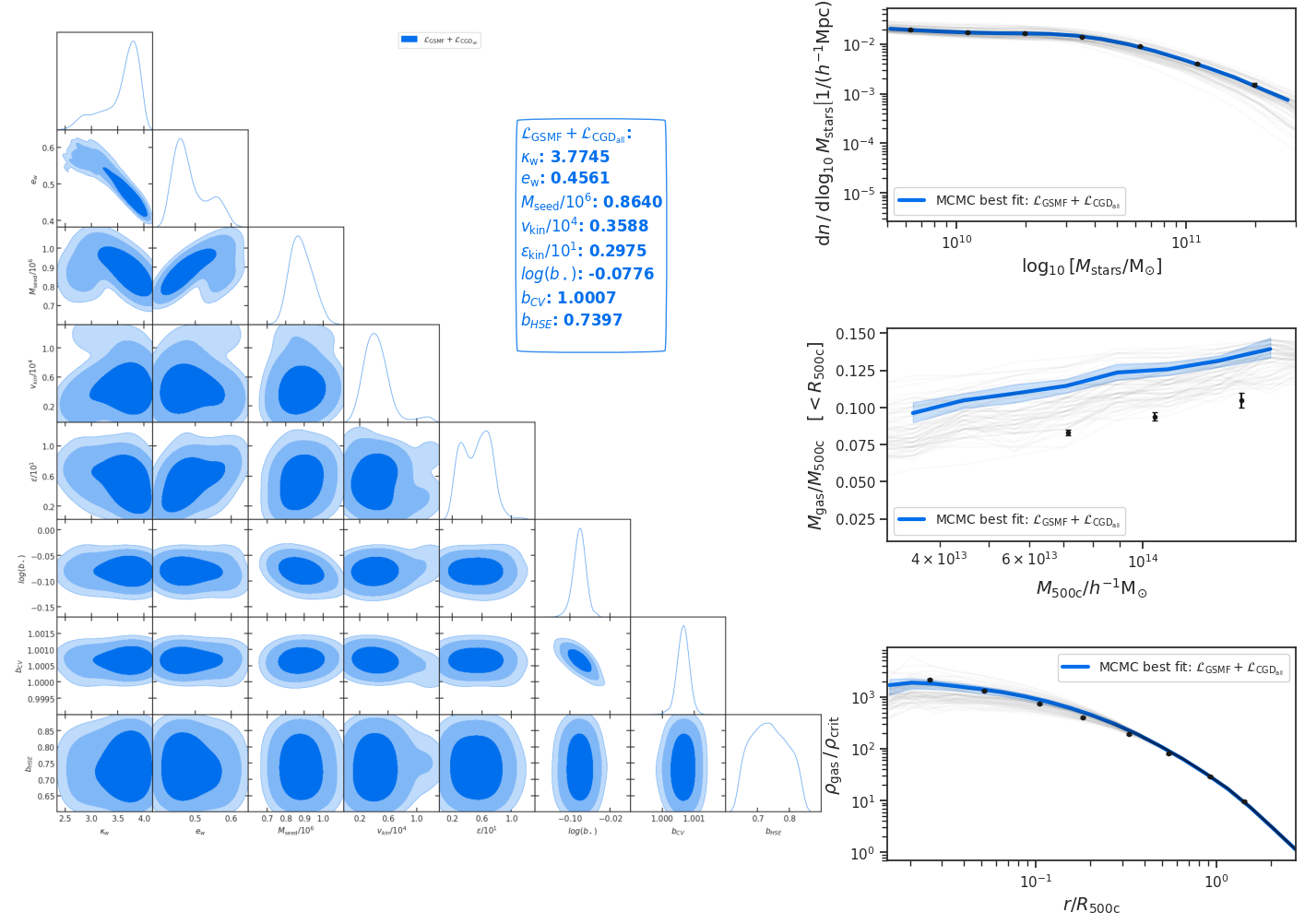}
    \caption{Bayesian posterior distribution of five subgrid parameters and three bias parameters. The forward model involves the joint likelihood of the $\rm{GSMF}$ and the $\mathrm{CGD}$ profiles. The right panels show the predictions at the best-fit parameters and the observational targets after applying bias corrections.}
    \label{fig:mcmc_b}
\end{figure*}

\textbf{The stellar mass bias} term ($b_*$) accounts for systematic uncertainties in the stellar mass determination from observations. Various systematic discrepancies exist between different observational studies, primarily stemming from the stellar population synthesis and dust correction models used. This mass-independent bias is modeled as $\log_{10}(M_{*,\text{obs}}) \rightarrow \log_{10}(M_{*,\text{obs}}) + \log_{10} b_*$, where a positive $b_*$ implies the observations underestimate the true stellar mass. In the Bayesian parameter calibration, we apply a lognormal prior on this parameter following \citet{Behroozi2019}: $\log_{10} b_* \sim \mathcal{N}(0, 0.14)$. This prior is informed by existing tensions between observed time-integrated star formation rates and observed stellar mass functions.

\textbf{The cosmic variance bias} ($b_{\text{CV}}$) accounts for the effect of limited survey volumes on number density measurements. As shown by \citet{Driver2010}, the error on $\rm{GSMF}$ due to cosmic variance can be 5-10\% for surveys like GAMA and SDSS, depending on the volume considered. Cosmic variance can bias the number density measurements because the survey may consist of slightly over- or under-dense regions. We model this as $\Phi_{\text{obs}} \rightarrow \Phi_{\text{obs}} + \log_{10}(b_{\text{CV}})$, where $\Phi_{\text{obs}}$ is the observed number density in the stellar mass function, and a positive $b_{\text{CV}}$ indicates that observations underestimate the true number density. For our mass range, we assume that this effect is also independent of mass. In the Bayesian inference, we impose a Gaussian prior $b_{\text{CV}} \sim \mathcal{N}(1, 0.06)$ based on the estimate that the error due to cosmic variance is about 6\% (following the discussions in \citet{Driver2022, kugel2023flamingo}). 

\textbf{The hydrostatic mass bias} ($b_{\text{HSE}}$) accounts for the well-documented underestimation of cluster masses when derived under the assumption of hydrostatic equilibrium \citep{Hoekstra2015, Eckert2016, Smith2016}. For X-ray observations, the density and temperature profiles fitted to the observations are used to measure the total mass, assuming the gas is in hydrostatic equilibrium (HSE). This approach is known to be biased low due to non-thermal pressure components and other departures from the HSE assumption. We define this bias as $\log_{10} M_{500c} = \log_{10} M_{500c,\text{HSE}} - \log_{10}(b_{\text{HSE}})$, where values $b_{\text{HSE}} < 1$ imply that hydrostatic mass measurements underestimate the true mass. We neglect the effect of hydrostatic bias on the gas fraction itself, as this effect is comparatively small \citep{McCarthy2017, Velliscig2014}. This is because both the total and gas mass increase with increasing $R_{500c}$, and the measured gas fraction will differ only at the level of the change in cumulative gas fraction between the true and biased $R_{500c}$.

Based on combined constraints from \citet{Eckert2016} and \citet{Hoekstra2015}, where the hydrostatic mass bias is estimated directly by comparing masses obtained from weak lensing and X-rays, we use the Gaussian prior $b_{\text{HSE}} \sim \mathcal{N}(0.74, 0.10)$. When fitting our model to observations, the hydrostatic bias primarily affects the $f_\mathrm{gas}$ measurements by shifting the inferred cluster masses, which is critical for properly comparing simulations to X-ray data.

In our Bayesian inference trial, we simultaneously fit for these 3 observational biases alongside the physical parameters of our model. The results from the MCMC run with included bias terms are shown in Figure~\ref{fig:mcmc_b}.

By accounting for both random and systematic errors, we reduce the risk of biasing our calibration due to observational systematics, resulting in more robust predictions for other observables. Similar to \citep{kugel2023flamingo}, our bias terms corresponding to stellar mass and cosmic variance are negligible (i.e., $\log_{10} b_* = -0.0776$ and $b_{\text{CV}} = 1.0007$), and the hydrostatic bias is measured to be $b_{\text{HSE}} = 0.7397$.

\section{Discrepancy factor}
\label{sec:appendix_disc}

We note an important caveat with respect to the $\mathrm{CGD}$ profile and present an illustrative exercise to quantify its sensitivity to simulation volume. Simulations in \autoref{sec:phase1} with side-length of $L = 128\,h^{-1}$Mpc result in a limited number of massive clusters ($\leq1$ cluster with a mass larger than $10^{14.5}$\,M$_\odot$ per simulation). For one of the data points in the experimental design, we ran an equivalent CRK-HACC simulation with $L = 256\,h^{-1}$Mpc using the same mass resolution as the smaller box. The larger box hosts $\sim 10$ clusters with masses larger than $10^{14.5}$\,M$_\odot$, which yields an improved estimate for $\mathrm{CGD}$. The resulting discrepancy term $\delta_0 (r/R_{500c})$ is then added to the extracted cluster gas densities for all the simulations as an approximate bias correction term. This zeroth-order correction in $\mathrm{CGD}$ ignores any subgrid parameter dependency and adds to the error source in the parameter constraints obtained from the Phase-1 emulators. We also note that this type of simulation-volume-dependent bias is limited to profile measurements and is not observed in $\rm{GSMF}$ or $f_\mathrm{gas}$. 

\begin{figure}
\includegraphics[width=0.45\textwidth]{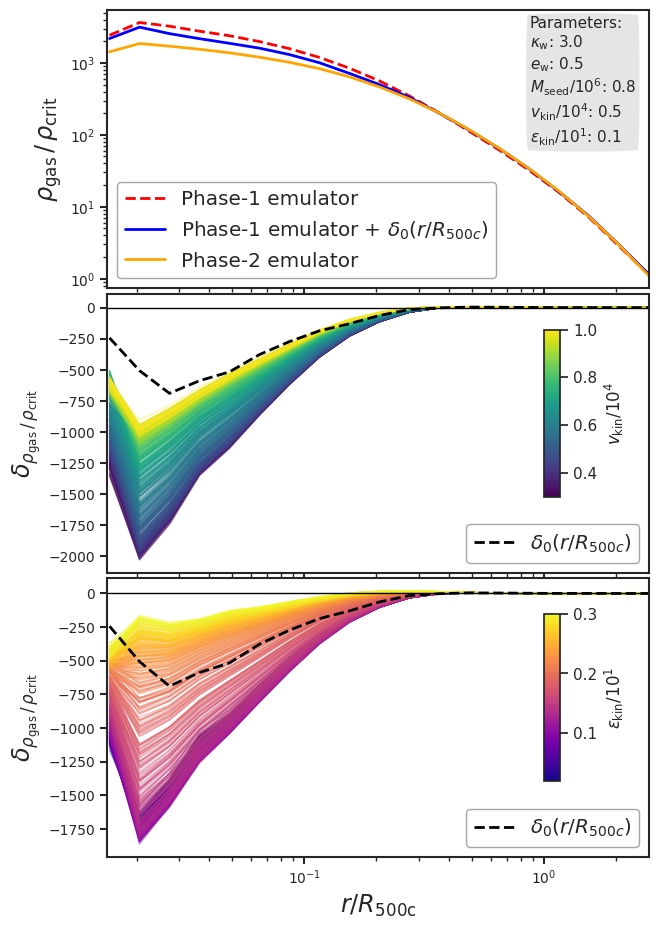}
\caption{Comparison of $\mathrm{CGD}$ values in Phase-1 and Phase-2 emulators. Top panel: For a fiducial value of sub-grid parameters, the red dashed line shows the Phase-1 emulator and the blue line shows the corrected $\mathrm{CGD}$ profiles with discrepancy term $\delta_0 (r/R_{\rm 500c})$ added. The orange line shows the emulator prediction from Phase-2. The middle and bottom panels show the discrepancy between Phase-2 and Phase-1 emulator estimates of $\mathrm{CGD}$ for varying kinetic feedback parameters.}
\label{fig:delta_phase2}
\end{figure}

We demonstrate the primary advantage of Phase-2 simulation ensemble and resulting emulators in \autoref{fig:delta_phase2}. We previously used a correction factor $\delta_0 (r/R_{\rm 500c})$ for correcting $\mathrm{CGD}$ profiles. Whereas when we consider the subgrid parameter dependency of the discrepancy between 2 simulation volumes, $\delta_{\rho_{gas} / \rho_{crit} } =  \mathrm{CGD}_{\rm all, Phase-2} - \mathrm{CGD}_{\rm all, Phase-1}$, we identify a clear high-order dependence with kinetic feedback parameters, as shown in the middle and bottom panels of \autoref{fig:delta_phase2}. Specifically, lower values of $v_\text{kin}$ and $\epsilon_\text{kin}$ display larger discrepancy (i.e., larger simulation volumes result in more suppressed $\mathrm{CGD}$ profiles). We also note that the parameter of the simulation pair used to compute the discrepancy is different from the fiducial parameters used in \autoref{fig:delta_phase2}, which results in $\delta_{\rho_{gas} / \rho_{crit} }$ values in middle panel to be lower than $\delta_0 (r/R_{\rm 500c})$. 

\begin{figure*}
    \includegraphics[width=1\textwidth]{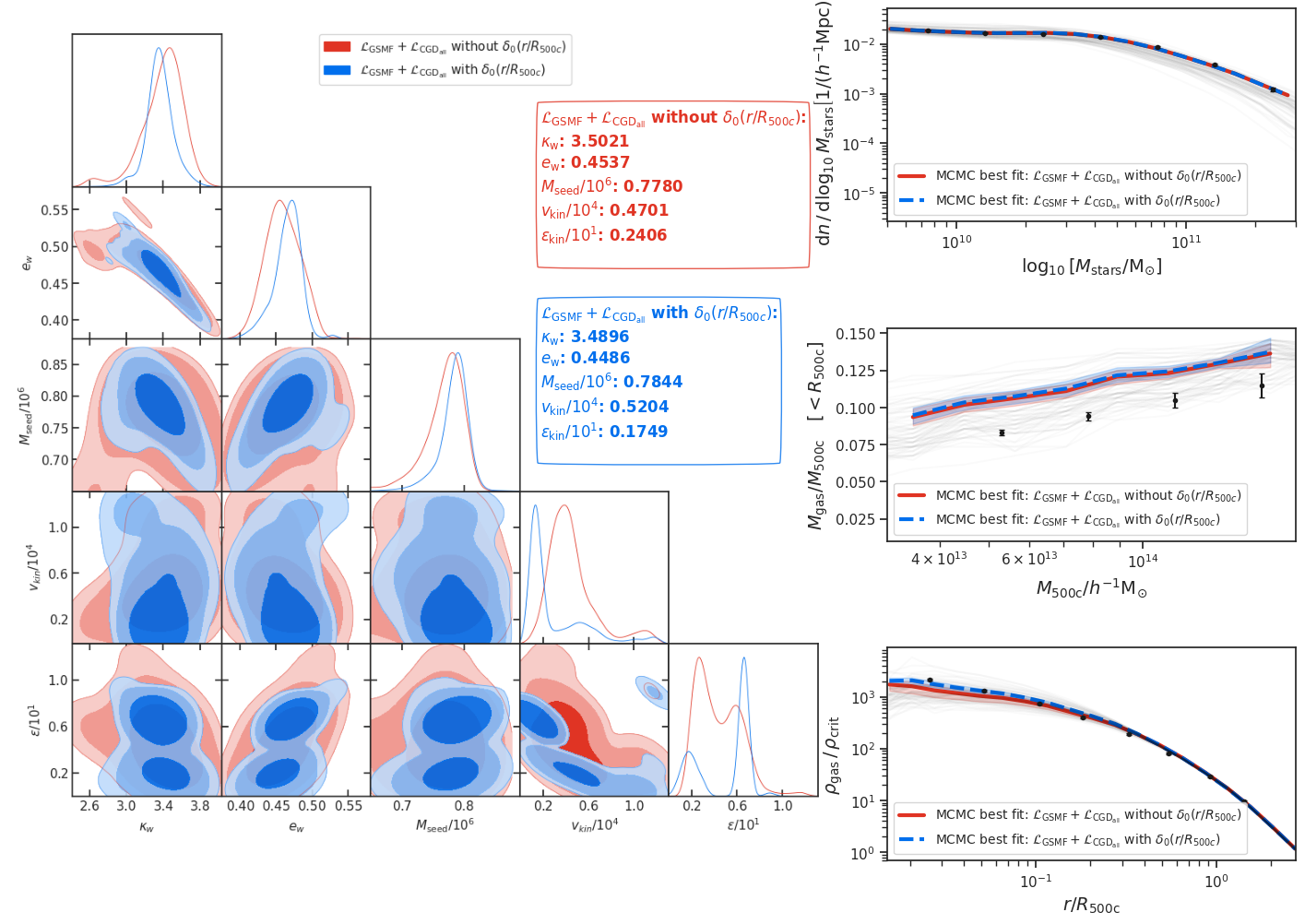}
    \caption{In Phase-1, The effect of $\mathrm{CGD}$ correction term $\delta_0 (r/R_{500c})$ is checked with MCMC runs. While the best-fit line for $\rm{GSMF}$ is not varied significantly, the $\mathrm{CGD}$ variation is noticeable.}
    \label{fig:mcmc_disc}
\end{figure*}

To further validate our results from \autoref{sec:calibration}, we perform experiments both with and without the bias correction factor $\delta_0 (r/R_{\rm 500c})$ applied to the circumgalactic density profiles. This correction addresses the limited cluster statistics inherent in our 128 Mpc/h simulation boxes, which yield only approximately 10 clusters. The inclusion of this correction produces two notable effects: The circumgalactic density profiles exhibit slightly elevated values at small radii, indicating denser gas concentrations in cluster cores, while the predicted gas fractions become even more discrepant with observations when not included in the calibration.

The discrepancy correction also impacts the optimal kinetic feedback parameters, reducing $\epsilon_\text{kin}$ from 2.4 to 1.8 and increasing $v_\text{kin}$ from $0.47 \times 10^4~\rm{km/s}$ to $0.52 \times 10^4~\rm{km/s}$. However, as shown in the posterior distributions in \autoref{fig:mcmc_disc}, including the discrepancy factor also reveals new modes within the low-kinetic feedback regime. Regardless, the inclusion of a subgrid-parameter independent discrepancy term does not resolve the fundamental tension: Simultaneously fitting both the gas fraction and circumgalactic density profiles with a single set of subgrid parameters is not possible with our subgrid physics models and target datasets. Instead, the bias correction reinforces the viability of a low-kinetic feedback regime that preferentially matches the circumgalactic density profiles at the expense of gas fraction accuracy.

We noted the discrepancy of $\mathrm{CGD}$ in the inner parts of the cluster due to the low cluster count in small-volume simulations. Using the single offset vector across all the subgrid parameter space is not ideal either. Hence, a deeper investigation involving $\mathrm{CGD}$ would require a new large-volume simulation suite, as carried out in \autoref{sec:phase2}.

\bibliographystyle{aasjournal}
\bibliography{biblio}

@ARTICLE{Moran2023,
       author = {{Moran}, Kelly R. and {Heitmann}, Katrin and {Lawrence}, Earl and {Habib}, Salman and {Bingham}, Derek and {Upadhye}, Amol and {Kwan}, Juliana and {Higdon}, David and {Payne}, Richard},
        title = "{The Mira-Titan Universe - IV. High-precision power spectrum emulation}",
      journal = {\mnras},
         year = 2023,
        month = apr,
       volume = {520},
       number = {3},
        pages = {3443-3458},
          doi = {10.1093/mnras/stac3452},
archivePrefix = {arXiv},
       eprint = {2207.12345},
 primaryClass = {astro-ph.CO},
       adsurl = {https://ui.adsabs.harvard.edu/abs/2023MNRAS.520.3443M}
}

@ARTICLE{Kwan2023,
       author = {{Kwan}, Juliana and {Saito}, Shun and {Leauthaud}, Alexie and {Heitmann}, Katrin and {Habib}, Salman and {Frontiere}, Nicholas and {Guo}, Hong and {Huang}, Song and {Pope}, Adrian and {Rodrigu{\'e}z-Torres}, Sergio},
        title = "{Galaxy Clustering in the Mira-Titan Universe. I. Emulators for the Redshift Space Galaxy Correlation Function and Galaxy-Galaxy Lensing}",
      journal = {\apj},
         year = 2023,
        month = jul,
       volume = {952},
       number = {1},
          eid = {80},
        pages = {80},
          doi = {10.3847/1538-4357/acd92f},
archivePrefix = {arXiv},
       eprint = {2302.12379},
 primaryClass = {astro-ph.CO},
adsurl = {https://ui.adsabs.harvard.edu/abs/2023ApJ...952...80K}
}

@ARTICLE{Crain2023,
       author = {{Crain}, Robert A. and {van de Voort}, Freeke},
        title = "{Hydrodynamical Simulations of the Galaxy Population: Enduring Successes and Outstanding Challenges}",
      journal = {\araa},
         year = 2023,
        month = aug,
       volume = {61},
        pages = {473-515},
          doi = {10.1146/annurev-astro-041923-043618},
archivePrefix = {arXiv},
       eprint = {2309.17075},
 primaryClass = {astro-ph.GA},
       adsurl = {https://ui.adsabs.harvard.edu/abs/2023ARA&A..61..473C}
}

@ARTICLE{Vogelsberger2020,
       author = {{Vogelsberger}, Mark and {Marinacci}, Federico and {Torrey}, Paul and {Puchwein}, Ewald},
        title = "{Cosmological simulations of galaxy formation}",
      journal = {Nature Reviews Physics},
         year = 2020,
        month = jan,
       volume = {2},
       number = {1},
        pages = {42-66},
          doi = {10.1038/s42254-019-0127-2},
archivePrefix = {arXiv},
       eprint = {1909.07976},
 primaryClass = {astro-ph.GA},
       adsurl = {https://ui.adsabs.harvard.edu/abs/2020NatRP...2...42V}
}

@ARTICLE{Valentini2025,
       author = {{Valentini}, Milena and {Dolag}, Klaus},
        title = "{Hydrodynamic methods and sub-resolution models for cosmological simulations}",
      journal = {arXiv e-prints},
         year = 2025,
        month = feb,
          eid = {arXiv:2502.06954},
        pages = {arXiv:2502.06954},
          doi = {10.48550/arXiv.2502.06954},
archivePrefix = {arXiv},
       eprint = {2502.06954},
 primaryClass = {astro-ph.CO},
       adsurl = {https://ui.adsabs.harvard.edu/abs/2025arXiv250206954V}
}

@article{rasmussen2009temperature,
  title={Temperature and abundance profiles of hot gas in galaxy groups--II. Implications for feedback and ICM enrichment},
  author={Rasmussen, Jesper and Ponman, Trevor J},
  journal={Monthly Notices of the Royal Astronomical Society},
  volume={399},
  number={1},
  pages={239--263},
  year={2009},
  publisher={Blackwell Publishing Ltd Oxford, UK}
}

@article{vikhlinin2009,
	author = {Vikhlinin, A and Burenin, RA and Ebeling, H and Forman, WR and Hornstrup, Allan and Jones, C and Kravtsov, AV and Murray, SS and Nagai, D and Quintana, H and others},
	journal = {The Astrophysical Journal},
	number = {2},
	pages = {1033},
	publisher = {IOP Publishing},
	title = {Chandra cluster cosmology project. II. Samples and X-ray data reduction},
	volume = {692},
	year = {2009}}

@article{heitmann2013coyote,
  title={The coyote universe extended: precision emulation of the matter power spectrum},
  author={Heitmann, Katrin and Lawrence, Earl and Kwan, Juliana and Habib, Salman and Higdon, David},
  journal={The Astrophysical Journal},
  volume={780},
  number={1},
  pages={111},
  year={2013},
  publisher={IOP Publishing}
}

@article{frontiere2025modeling,
  title={Modeling Galaxy Formation in Cosmological Simulations with CRK-HACC},
  author={Frontiere, Nicholas and Emberson, JD and Buehlmann, Michael and Habib, Salman and Heitmann, Katrin and Ramachandra, Nesar and Faucher-Gigu{\`e}re, Claude-Andr{\'e}},
  journal={arXiv preprint arXiv:2511.21921},
  year={2025}
}

@inproceedings{frontiere2025cosmological,
  title={Cosmological Hydrodynamics at Exascale: A Trillion-Particle Leap in Capability},
  author={Frontiere, Nicholas and Emberson, JD and Buehlmann, Michael and Rangel, Esteban M and Habib, Salman and Heitmann, Katrin and Larsen, Patricia and Morozov, Vitali and Pope, Adrian and Faucher-Gigu{\`e}re, Claude-Andr{\'e} and others},
  booktitle={Proceedings of the International Conference for High Performance Computing, Networking, Storage and Analysis},
  pages={25--35},
  year={2025}
}

@article{mulroy2019locuss,
  title={LoCuSS: scaling relations between galaxy cluster mass, gas, and stellar content},
  author={Mulroy, Sarah L and Farahi, Arya and Evrard, August E and Smith, Graham P and Finoguenov, Alexis and O’Donnell, Christine and Marrone, Daniel P and Abdulla, Zubair and Bourdin, Herv{\'e} and Carlstrom, John E and others},
  journal={Monthly Notices of the Royal Astronomical Society},
  volume={484},
  number={1},
  pages={60--80},
  year={2019},
  publisher={Oxford University Press}
}

@article{hoekstra2015canadian,
  title={The Canadian Cluster Comparison Project: detailed study of systematics and updated weak lensing masses},
  author={Hoekstra, Henk and Herbonnet, Ricardo and Muzzin, Adam and Babul, Arif and Mahdavi, Andi and Viola, Massimo and Cacciato, Marcello},
  journal={Monthly Notices of the Royal Astronomical Society},
  volume={449},
  number={1},
  pages={685--714},
  year={2015},
  publisher={Oxford University Press}
}

@article{pearson2017galaxy,
  title={Galaxy And Mass Assembly: search for a population of high-entropy galaxy groups},
  author={Pearson, Richard J and Ponman, Trevor J and Norberg, Peder and Robotham, Aaron SG and Babul, Arif and Bower, Richard G and McCarthy, Ian G and Brough, Sarah and Driver, Simon P and Pimbblet, Kevin},
  journal={Monthly Notices of the Royal Astronomical Society},
  volume={469},
  number={3},
  pages={3489--3504},
  year={2017},
  publisher={Oxford University Press}
}

@article{lovisari2020x,
  title={X-ray scaling relations for a representative sample of Planck-selected clusters observed with XMM-Newton},
  author={Lovisari, Lorenzo and Schellenberger, Gerrit and Sereno, Mauro and Ettori, Stefano and Pratt, Gabriel W and Forman, William R and Jones, Christine and Andrade-Santos, Felipe and Randall, Scott and Kraft, Ralph},
  journal={The Astrophysical Journal},
  volume={892},
  number={2},
  pages={102},
  year={2020},
  publisher={IOP Publishing}
}

@article{lovisari2015scaling,
  title={Scaling properties of a complete X-ray selected galaxy group sample},
  author={Lovisari, Lorenzo and Reiprich, TH and Schellenberger, Gerrit},
  journal={Astronomy \& Astrophysics},
  volume={573},
  pages={A118},
  year={2015},
  publisher={EDP Sciences}
}

@article{sanderson2013baryon,
  title={The baryon budget on the galaxy group/cluster boundary},
  author={Sanderson, Alastair JR and O'Sullivan, Ewan and Ponman, Trevor J and Gonzalez, Anthony H and Sivanandam, Suresh and Zabludoff, Ann I and Zaritsky, Dennis},
  journal={Monthly Notices of the Royal Astronomical Society},
  volume={429},
  number={4},
  pages={3288--3304},
  year={2013},
  publisher={Oxford University Press}
}

@article{gonzalez2013galaxy,
  title={Galaxy cluster baryon fractions revisited},
  author={Gonzalez, Anthony H and Sivanandam, Suresh and Zabludoff, Ann I and Zaritsky, Dennis},
  journal={The Astrophysical Journal},
  volume={778},
  number={1},
  pages={14},
  year={2013},
  publisher={IOP Publishing}
}

@article{lagana2013comprehensive,
  title={A comprehensive picture of baryons in groups and clusters of galaxies},
  author={Lagan{\'a}, TF and Martinet, N and Durret, F and Neto, GB Lima and Maughan, B and Zhang, Y-Y},
  journal={Astronomy \& Astrophysics},
  volume={555},
  pages={A66},
  year={2013},
  publisher={EDP Sciences}
}

@article{lin2011baryon,
  title={Baryon Content of Massive Galaxy Clusters at z= 0--0.6},
  author={Lin, Yen-Ting and Stanford, S Adam and Eisenhardt, Peter RM and Vikhlinin, Alexey and Maughan, Ben J and Kravtsov, Andrey},
  journal={The Astrophysical Journal Letters},
  volume={745},
  number={1},
  pages={L3},
  year={2011},
  publisher={IOP Publishing}
}

@article{sun2009chandra,
  title={Chandra studies of the X-ray gas properties of galaxy groups},
  author={Sun, M and Voit, GM and Donahue, M and Jones, C and Forman, W and Vikhlinin, A},
  journal={The Astrophysical Journal},
  volume={693},
  number={2},
  pages={1142},
  year={2009},
  publisher={IOP Publishing}
}

@article{pratt2010gas,
  title={Gas entropy in a representative sample of nearby X-ray galaxy clusters (REXCESS): relationship to gas mass fraction},
  author={Pratt, GW and Arnaud, M and Piffaretti, R and B{\"o}hringer, H and Ponman, TJ and Croston, Judith H and Voit, GM and Borgani, Stefano and Bower, RG},
  journal={Astronomy \& Astrophysics},
  volume={511},
  pages={A85},
  year={2010},
  publisher={EDP Sciences}
}

@article{maughan2008images,
  title={Images, structural properties, and metal abundances of galaxy clusters observed with Chandra ACIS-I at 0.1< z< 1.3},
  author={Maughan, BJ and Jones, C and Forman, W and Van Speybroeck, L},
  journal={The Astrophysical Journal Supplement Series},
  volume={174},
  number={1},
  pages={117},
  year={2008},
  publisher={IOP Publishing}
}

@INPROCEEDINGS{1996kddm.conf..226E,
       author = {{Ester}, Martin and {Kriegel}, Hans-Peter and {Sander}, J{\"o}rg and {Xu}, Xiaowei},
        title = "{A Density-Based Algorithm for Discovering Clusters in Large Spatial Databases with Noise}",
    booktitle = {Second International Conference on Knowledge Discovery and Data Mining (KDD'96). Proceedings of a conference held August 2-4},
         year = 1996,
       editor = {{Pfitzner}, D.~W. and {Salmon}, J.~K.},
        month = jan,
        pages = {226-331},
       adsurl = {https://ui.adsabs.harvard.edu/abs/1996kddm.conf..226E}
}

@ARTICLE{1993ApJ...416....1K,
       author = {{Klypin}, Anatoly and {Holtzman}, Jon and {Primack}, Joel and {Regos}, Eniko},
        title = "{Structure Formation with Cold plus Hot Dark Matter}",
      journal = {\apj},
         year = 1993,
        month = oct,
       volume = {416},
        pages = {1},
          doi = {10.1086/173210},
archivePrefix = {arXiv},
       eprint = {astro-ph/9305011},
 primaryClass = {astro-ph},
       adsurl = {https://ui.adsabs.harvard.edu/abs/1993ApJ...416....1K}
}

@ARTICLE{1985ApJ...292..371D,
       author = {{Davis}, M. and {Efstathiou}, G. and {Frenk}, C.~S. and {White}, S.~D.~M.},
        title = "{The evolution of large-scale structure in a universe dominated by cold dark matter}",
      journal = {\apj},
         year = 1985,
        month = may,
       volume = {292},
        pages = {371-394},
          doi = {10.1086/163168},
       adsurl = {https://ui.adsabs.harvard.edu/abs/1985ApJ...292..371D}
}

@ARTICLE{1996MNRAS.278..488Z,
       author = {{Zhao}, Hongsheng},
        title = "{Analytical models for galactic nuclei}",
      journal = {\mnras},
         year = 1996,
        month = jan,
       volume = {278},
       number = {2},
        pages = {488-496},
          doi = {10.1093/mnras/278.2.488},
archivePrefix = {arXiv},
       eprint = {astro-ph/9509122},
 primaryClass = {astro-ph},
       adsurl = {https://ui.adsabs.harvard.edu/abs/1996MNRAS.278..488Z}
}

@article{bigwood2025case,
  title={The case for large-scale AGN feedback in galaxy formation simulations: insights from XFABLE},
  author={Bigwood, Leah and Bourne, Martin A and Irsic, Vid and Amon, Alexandra and Sijacki, Debora},
  journal={arXiv preprint arXiv:2501.16983},
  year={2025}
}

@ARTICLE{2009ApJ...696..620C,
       author = {{Conroy}, Charlie and {Wechsler}, Risa H.},
        title = "{Connecting Galaxies, Halos, and Star Formation Rates Across Cosmic Time}",
      journal = {\apj},
         year = 2009,
        month = may,
       volume = {696},
       number = {1},
        pages = {620-635},
          doi = {10.1088/0004-637X/696/1/620},
archivePrefix = {arXiv},
       eprint = {0805.3346},
 primaryClass = {astro-ph},
       adsurl = {https://ui.adsabs.harvard.edu/abs/2009ApJ...696..620C}
}

@article{wechsler2018connection,
  title={The connection between galaxies and their dark matter halos},
  author={Wechsler, Risa H and Tinker, Jeremy L},
  journal={Annual Review of Astronomy and Astrophysics},
  volume={56},
  number={1},
  pages={435--487},
  year={2018},
  publisher={Annual Reviews}
}

@article{madau2014cosmic,
  title={Cosmic star-formation history},
  author={Madau, Piero and Dickinson, Mark},
  journal={Annual Review of Astronomy and Astrophysics},
  volume={52},
  number={1},
  pages={415--486},
  year={2014},
  publisher={Annual Reviews}
}

@article{bondi1944,
	author = {Bondi, Hermann and Hoyle, Fred},
	journal = {Monthly Notices of the Royal Astronomical Society},
	number = {5},
	pages = {273--282},
	publisher = {Oxford Academic},
	title = {{On the mechanism of accretion by stars}},
	volume = {104},
	year = {1944}}

@inproceedings{hoyle1939,
	author = {Hoyle, Fred and Lyttleton, Raymond A},
	booktitle = {Mathematical Proceedings of the Cambridge Philosophical Society},
	month = {July},
	organization = {Cambridge University Press},
	pages = {405--415},
	title = {{The effect of interstellar matter on climatic variation}},
	volume = {35},
	year = {1939}}

@ARTICLE{1999ApJS..123....3L,
       author = {{Leitherer}, Claus and {Schaerer}, Daniel and {Goldader}, Jeffrey D. and {Delgado}, Rosa M. Gonz{\'a}lez and {Robert}, Carmelle and {Kune}, Denis Foo and {de Mello}, Du{\'\i}lia F. and {Devost}, Daniel and {Heckman}, Timothy M.},
        title = "{Starburst99: Synthesis Models for Galaxies with Active Star Formation}",
      journal = {\apjs},
         year = 1999,
        month = jul,
       volume = {123},
       number = {1},
        pages = {3-40},
          doi = {10.1086/313233},
archivePrefix = {arXiv},
       eprint = {astro-ph/9902334},
 primaryClass = {astro-ph},
       adsurl = {https://ui.adsabs.harvard.edu/abs/1999ApJS..123....3L}
}

@ARTICLE{2017MNRAS.467.4739K,
       author = {{Kaviraj}, S. and {Laigle}, C. and {Kimm}, T. and {Devriendt}, J.~E.~G. and {Dubois}, Y. and {Pichon}, C. and {Slyz}, A. and {Chisari}, E. and {Peirani}, S.},
        title = "{The Horizon-AGN simulation: evolution of galaxy properties over cosmic time}",
      journal = {\mnras},
         year = 2017,
        month = jun,
       volume = {467},
       number = {4},
        pages = {4739-4752},
          doi = {10.1093/mnras/stx126},
archivePrefix = {arXiv},
       eprint = {1605.09379},
 primaryClass = {astro-ph.GA},
       adsurl = {https://ui.adsabs.harvard.edu/abs/2017MNRAS.467.4739K}
}

@ARTICLE{2019MNRAS.486.2827D,
       author = {{Dav{\'e}}, Romeel and {Angl{\'e}s-Alc{\'a}zar}, Daniel and {Narayanan}, Desika and {Li}, Qi and {Rafieferantsoa}, Mika H. and {Appleby}, Sarah},
        title = "{SIMBA: Cosmological simulations with black hole growth and feedback}",
      journal = {\mnras},
         year = 2019,
        month = jun,
       volume = {486},
       number = {2},
        pages = {2827-2849},
          doi = {10.1093/mnras/stz937},
archivePrefix = {arXiv},
       eprint = {1901.10203},
 primaryClass = {astro-ph.GA},
       adsurl = {https://ui.adsabs.harvard.edu/abs/2019MNRAS.486.2827D}
}

@ARTICLE{Behroozi2019,
       author = {{Behroozi}, Peter and {Wechsler}, Risa H. and {Hearin}, Andrew P. and {Conroy}, Charlie},
        title = "{UNIVERSEMACHINE: The correlation between galaxy growth and dark matter halo assembly from z = 0-10}",
      journal = {\mnras},
         year = 2019,
        month = sep,
       volume = {488},
       number = {3},
        pages = {3143-3194},
          doi = {10.1093/mnras/stz1182},
archivePrefix = {arXiv},
       eprint = {1806.07893},
 primaryClass = {astro-ph.GA},
       adsurl = {https://ui.adsabs.harvard.edu/abs/2019MNRAS.488.3143B}
}

@ARTICLE{2021ApJS..252...19H,
       author = {{Heitmann}, Katrin and {Frontiere}, Nicholas and {Rangel}, Esteban and {Larsen}, Patricia and {Pope}, Adrian and {Sultan}, Imran and {Uram}, Thomas and {Habib}, Salman and {Finkel}, Hal and {Korytov}, Danila and {Kovacs}, Eve and {Rizzi}, Silvio and {Insley}, Joe and {Knowles}, Janet Y.~K.},
        title = "{The Last Journey. I. An Extreme-scale Simulation on the Mira Supercomputer}",
      journal = {\apjs},
         year = 2021,
        month = feb,
       volume = {252},
       number = {2},
          eid = {19},
        pages = {19},
          doi = {10.3847/1538-4365/abcc67},
archivePrefix = {arXiv},
       eprint = {2006.01697},
 primaryClass = {astro-ph.CO},
       adsurl = {https://ui.adsabs.harvard.edu/abs/2021ApJS..252...19H}
}

@article{faucher2020cosmic,
  title={A cosmic UV/X-ray background model update},
  author={Faucher-Giguere, Claude-Andr{\'e}},
  journal={Monthly Notices of the Royal Astronomical Society},
  volume={493},
  number={2},
  pages={1614--1632},
  year={2020},
  publisher={Oxford University Press}
}

@article{potter2017pkdgrav3,
  title={PKDGRAV3: beyond trillion particle cosmological simulations for the next era of galaxy surveys},
  author={Potter, Douglas and Stadel, Joachim and Teyssier, Romain},
  journal={Computational Astrophysics and Cosmology},
  volume={4},
  number={1},
  pages={2},
  year={2017},
  publisher={Springer}
}

@ARTICLE{Pfeifer2020bahamas,
       author = {{Pfeifer}, Simon and {McCarthy}, Ian G. and {Stafford}, Sam G. and {Brown}, Shaun T. and {Font}, Andreea S. and {Kwan}, Juliana and {Salcido}, Jaime and {Schaye}, Joop},
        title = "{The BAHAMAS project: effects of dynamical dark energy on large-scale structure}",
      journal = {\mnras},
         year = 2020,
        month = oct,
       volume = {498},
       number = {2},
        pages = {1576-1592},
          doi = {10.1093/mnras/staa2240},
archivePrefix = {arXiv},
       eprint = {2004.07670},
 primaryClass = {astro-ph.CO},
       adsurl = {https://ui.adsabs.harvard.edu/abs/2020MNRAS.498.1576P}
}

@ARTICLE{Pakmor2023MillenniumTNG,
       author = {{Pakmor}, R{\"u}diger and {Springel}, Volker and {Coles}, Jonathan P. and {Guillet}, Thomas and {Pfrommer}, Christoph and {Bose}, Sownak and {Barrera}, Monica and {Delgado}, Ana Maria and {Ferlito}, Fulvio and {Frenk}, Carlos and {Hadzhiyska}, Boryana and {Hern{\'a}ndez-Aguayo}, C{\'e}sar and {Hernquist}, Lars and {Kannan}, Rahul and {White}, Simon D.~M.},
        title = "{The MillenniumTNG Project: the hydrodynamical full physics simulation and a first look at its galaxy clusters}",
      journal = {\mnras},
         year = 2023,
        month = sep,
       volume = {524},
       number = {2},
        pages = {2539-2555},
          doi = {10.1093/mnras/stac3620},
archivePrefix = {arXiv},
       eprint = {2210.10060},
 primaryClass = {astro-ph.CO},
       adsurl = {https://ui.adsabs.harvard.edu/abs/2023MNRAS.524.2539P}
}

@ARTICLE{Villaescusa-Navarro2021camels,
       author = {{Villaescusa-Navarro}, Francisco and {Angl{\'e}s-Alc{\'a}zar}, Daniel and {Genel}, Shy and {Spergel}, David N. and {Somerville}, Rachel S. and {Dave}, Romeel and {Pillepich}, Annalisa and {Hernquist}, Lars and {Nelson}, Dylan and {Torrey}, Paul and {Narayanan}, Desika and {Li}, Yin and {Philcox}, Oliver and {La Torre}, Valentina and {Maria Delgado}, Ana and {Ho}, Shirley and {Hassan}, Sultan and {Burkhart}, Blakesley and {Wadekar}, Digvijay and {Battaglia}, Nicholas and {Contardo}, Gabriella and {Bryan}, Greg L.},
        title = "{The CAMELS Project: Cosmology and Astrophysics with Machine-learning Simulations}",
      journal = {\apj},
         year = 2021,
        month = jul,
       volume = {915},
       number = {1},
          eid = {71},
        pages = {71},
          doi = {10.3847/1538-4357/abf7ba},
archivePrefix = {arXiv},
       eprint = {2010.00619},
 primaryClass = {astro-ph.CO},
       adsurl = {https://ui.adsabs.harvard.edu/abs/2021ApJ...915...71V}
}

@article{schaye2023flamingo,
  title={The FLAMINGO project: cosmological hydrodynamical simulations for large-scale structure and galaxy cluster surveys},
  author={Schaye, Joop and Kugel, Roi and Schaller, Matthieu and Helly, John C and Braspenning, Joey and Elbers, Willem and McCarthy, Ian G and van Daalen, Marcel P and Vandenbroucke, Bert and Frenk, Carlos S and others},
  journal={Monthly Notices of the Royal Astronomical Society},
  volume={526},
  number={4},
  pages={4978--5020},
  year={2023},
  publisher={Oxford University Press}
}

@ARTICLE{Driver2022,
       author = {{Driver}, Simon P. and {Bellstedt}, Sabine and {Robotham}, Aaron S.~G. and {Baldry}, Ivan K. and {Davies}, Luke J. and {Liske}, Jochen and {Obreschkow}, Danail and {Taylor}, Edward N. and {Wright}, Angus H. and {Alpaslan}, Mehmet and {Bamford}, Steven P. and {Bauer}, Amanda E. and {Bland-Hawthorn}, Joss and {Bilicki}, Maciej and {Bravo}, Mat{\'\i}as and {Brough}, Sarah and {Casura}, Sarah and {Cluver}, Michelle E. and {Colless}, Matthew and {Conselice}, Christopher J. and {Croom}, Scott M. and {de Jong}, Jelte and {D'Eugenio}, Franceso and {De Propris}, Roberto and {Dogruel}, Burak and {Drinkwater}, Michael J. and {Dvornik}, Andrej and {Farrow}, Daniel J. and {Frenk}, Carlos S. and {Giblin}, Benjamin and {Graham}, Alister W. and {Grootes}, Meiert W. and {Gunawardhana}, Madusha L.~P. and {Hashemizadeh}, Abdolhosein and {H{\"a}u{\ss}ler}, Boris and {Heymans}, Catherine and {Hildebrandt}, Hendrik and {Holwerda}, Benne W. and {Hopkins}, Andrew M. and {Jarrett}, Tom H. and {Heath Jones}, D. and {Kelvin}, Lee S. and {Koushan}, Soheil and {Kuijken}, Konrad and {Lara-L{\'o}pez}, Maritza A. and {Lange}, Rebecca and {L{\'o}pez-S{\'a}nchez}, {\'A}ngel R. and {Loveday}, Jon and {Mahajan}, Smriti and {Meyer}, Martin and {Moffett}, Amanda J. and {Napolitano}, Nicola R. and {Norberg}, Peder and {Owers}, Matt S. and {Radovich}, Mario and {Raouf}, Mojtaba and {Peacock}, John A. and {Phillipps}, Steven and {Pimbblet}, Kevin A. and {Popescu}, Cristina and {Said}, Khaled and {Sansom}, Anne E. and {Seibert}, Mark and {Sutherland}, Will J. and {Thorne}, Jessica E. and {Tuffs}, Richard J. and {Turner}, Ryan and {van der Wel}, Arjen and {van Kampen}, Eelco and {Wilkins}, Steve M.},
        title = "{Galaxy And Mass Assembly (GAMA): Data Release 4 and the z < 0.1 total and z < 0.08 morphological galaxy stellar mass functions}",
      journal = {\mnras},
         year = 2022,
        month = jun,
       volume = {513},
       number = {1},
        pages = {439-467},
          doi = {10.1093/mnras/stac472},
archivePrefix = {arXiv},
       eprint = {2203.08539},
 primaryClass = {astro-ph.GA},
       adsurl = {https://ui.adsabs.harvard.edu/abs/2022MNRAS.513..439D}
}

@ARTICLE{Velliscig2014,
       author = {{Velliscig}, Marco and {van Daalen}, Marcel P. and {Schaye}, Joop and {McCarthy}, Ian G. and {Cacciato}, Marcello and {Le Brun}, Amandine M.~C. and {Dalla Vecchia}, Claudio},
        title = "{The impact of galaxy formation on the total mass, mass profile and abundance of haloes}",
      journal = {\mnras},
         year = 2014,
        month = aug,
       volume = {442},
       number = {3},
        pages = {2641-2658},
          doi = {10.1093/mnras/stu1044},
archivePrefix = {arXiv},
       eprint = {1402.4461},
 primaryClass = {astro-ph.CO},
       adsurl = {https://ui.adsabs.harvard.edu/abs/2014MNRAS.442.2641V}
}

@ARTICLE{McCarthy2017,
       author = {{McCarthy}, Ian G. and {Schaye}, Joop and {Bird}, Simeon and {Le Brun}, Amandine M.~C.},
        title = "{The BAHAMAS project: calibrated hydrodynamical simulations for large-scale structure cosmology}",
      journal = {\mnras},
         year = 2017,
        month = mar,
       volume = {465},
       number = {3},
        pages = {2936-2965},
          doi = {10.1093/mnras/stw2792},
archivePrefix = {arXiv},
       eprint = {1603.02702},
 primaryClass = {astro-ph.CO},
       adsurl = {https://ui.adsabs.harvard.edu/abs/2017MNRAS.465.2936M}
}

@ARTICLE{Smith2016,
       author = {{Smith}, G.~P. and {Mazzotta}, P. and {Okabe}, N. and {Ziparo}, F. and {Mulroy}, S.~L. and {Babul}, A. and {Finoguenov}, A. and {McCarthy}, I.~G. and {Lieu}, M. and {Bah{\'e}}, Y.~M. and {Bourdin}, H. and {Evrard}, A.~E. and {Futamase}, T. and {Haines}, C.~P. and {Jauzac}, M. and {Marrone}, D.~P. and {Martino}, R. and {May}, P.~E. and {Taylor}, J.~E. and {Umetsu}, K.},
        title = "{LoCuSS: Testing hydrostatic equilibrium in galaxy clusters}",
      journal = {\mnras},
         year = 2016,
        month = feb,
       volume = {456},
       number = {1},
        pages = {L74-L78},
          doi = {10.1093/mnrasl/slv175},
archivePrefix = {arXiv},
       eprint = {1511.01919},
 primaryClass = {astro-ph.CO},
       adsurl = {https://ui.adsabs.harvard.edu/abs/2016MNRAS.456L..74S}
}

@ARTICLE{Eckert2016,
       author = {{Eckert}, D. and {Ettori}, S. and {Coupon}, J. and {Gastaldello}, F. and {Pierre}, M. and {Melin}, J. -B. and {Le Brun}, A.~M.~C. and {McCarthy}, I.~G. and {Adami}, C. and {Chiappetti}, L. and {Faccioli}, L. and {Giles}, P. and {Lavoie}, S. and {Lef{\`e}vre}, J.~P. and {Lieu}, M. and {Mantz}, A. and {Maughan}, B. and {McGee}, S. and {Pacaud}, F. and {Paltani}, S. and {Sadibekova}, T. and {Smith}, G.~P. and {Ziparo}, F.},
        title = "{The XXL Survey. XIII. Baryon content of the bright cluster sample}",
      journal = {\aap},
         year = 2016,
        month = jun,
       volume = {592},
          eid = {A12},
        pages = {A12},
          doi = {10.1051/0004-6361/201527293},
archivePrefix = {arXiv},
       eprint = {1512.03814},
 primaryClass = {astro-ph.CO},
       adsurl = {https://ui.adsabs.harvard.edu/abs/2016A&A...592A..12E}
}

@ARTICLE{Hoekstra2015,
       author = {{Hoekstra}, Henk and {Herbonnet}, Ricardo and {Muzzin}, Adam and {Babul}, Arif and {Mahdavi}, Andi and {Viola}, Massimo and {Cacciato}, Marcello},
        title = "{The Canadian Cluster Comparison Project: detailed study of systematics and updated weak lensing masses}",
      journal = {\mnras},
         year = 2015,
        month = may,
       volume = {449},
       number = {1},
        pages = {685-714},
          doi = {10.1093/mnras/stv275},
archivePrefix = {arXiv},
       eprint = {1502.01883},
 primaryClass = {astro-ph.CO},
       adsurl = {https://ui.adsabs.harvard.edu/abs/2015MNRAS.449..685H}
}

@ARTICLE{Driver2010,
       author = {{Driver}, Simon P. and {Robotham}, Aaron S.~G.},
        title = "{Quantifying cosmic variance}",
      journal = {\mnras},
         year = 2010,
        month = oct,
       volume = {407},
       number = {4},
        pages = {2131-2140},
          doi = {10.1111/j.1365-2966.2010.17028.x},
archivePrefix = {arXiv},
       eprint = {1005.2538},
 primaryClass = {astro-ph.CO},
       adsurl = {https://ui.adsabs.harvard.edu/abs/2010MNRAS.407.2131D}
}

@article{smith2003stable,
  title={Stable clustering, the halo model and non-linear cosmological power spectra},
  author={Smith, Robert E and Peacock, J Aꎬ and Jenkins, Aꎬ and White, SDM and Frenk, CS and Pearce, FR and Thomas, Peter A and Efstathiou, G and Couchman, HMP},
  journal={Monthly Notices of the Royal Astronomical Society},
  volume={341},
  number={4},
  pages={1311--1332},
  year={2003},
  publisher={Blackwell Science Ltd Oxford, UK}
}

@article{takahashi2012revising,
  title={Revising the halofit model for the nonlinear matter power spectrum},
  author={Takahashi, Ryuichi and Sato, Masanori and Nishimichi, Takahiro and Taruya, Atsushi and Oguri, Masamune},
  journal={The Astrophysical Journal},
  volume={761},
  number={2},
  pages={152},
  year={2012},
  publisher={IOP Publishing}
}

@article{heitmann2019outer,
  title={The outer rim simulation: A path to many-core supercomputers},
  author={Heitmann, Katrin and Finkel, Hal and Pope, Adrian and Morozov, Vitali and Frontiere, Nicholas and Habib, Salman and Rangel, Esteban and Uram, Thomas and Korytov, Danila and Child, Hillary and others},
  journal={The Astrophysical Journal Supplement Series},
  volume={245},
  number={1},
  pages={16},
  year={2019},
  publisher={IOP Publishing}
}

@article{frontiere2017crksph,
  title={CRKSPH--a conservative reproducing kernel smoothed particle hydrodynamics scheme},
  author={Frontiere, Nicholas and Raskin, Cody D and Owen, J Michael},
  journal={Journal of Computational Physics},
  volume={332},
  pages={160--209},
  year={2017},
  publisher={Elsevier}
}

@article{frontiere2023simulating,
  title={Simulating hydrodynamics in cosmology with crk-hacc},
  author={Frontiere, Nicholas and Emberson, JD and Buehlmann, Michael and Adamo, Joseph and Habib, Salman and Heitmann, Katrin and Faucher-Gigu{\`e}re, Claude-Andr{\'e}},
  journal={The Astrophysical Journal Supplement Series},
  volume={264},
  number={2},
  pages={34},
  year={2023},
  publisher={IOP Publishing}
}

@article{ishiyama2021uchuu,
  title={The Uchuu simulations: Data Release 1 and dark matter halo concentrations},
  author={Ishiyama, Tomoaki and Prada, Francisco and Klypin, Anatoly A and Sinha, Manodeep and Metcalf, R Benton and Jullo, Eric and Altieri, Bruno and Cora, Sof{\'\i}a A and Croton, Darren and de La Torre, Sylvain and others},
  journal={Monthly Notices of the Royal Astronomical Society},
  volume={506},
  number={3},
  pages={4210--4231},
  year={2021},
  publisher={Oxford University Press}
}

@article{frontiere2022farpoint,
  title={Farpoint: A High-resolution Cosmology Simulation at the Gigaparsec Scale},
  author={Frontiere, Nicholas and Heitmann, Katrin and Rangel, Esteban and Larsen, Patricia and Pope, Adrian and Sultan, Imran and Uram, Thomas and Habib, Salman and Rizzi, Silvio and Insley, Joe and others},
  journal={The Astrophysical Journal Supplement Series},
  volume={259},
  number={1},
  pages={15},
  year={2022},
  publisher={IOP Publishing}
}

@article{weinberger2016simulating,
  title={Simulating galaxy formation with black hole driven thermal and kinetic feedback},
  author={Weinberger, Rainer and Springel, Volker and Hernquist, Lars and Pillepich, Annalisa and Marinacci, Federico and Pakmor, R{\"u}diger and Nelson, Dylan and Genel, Shy and Vogelsberger, Mark and Naiman, Jill and others},
  journal={Monthly Notices of the Royal Astronomical Society},
  volume={465},
  number={3},
  pages={3291--3308},
  year={2016},
  publisher={The Royal Astronomical Society}
}

@article{wiersma2009effect,
  title={The effect of photoionization on the cooling rates of enriched, astrophysical plasmas},
  author={Wiersma, Robert PC and Schaye, Joop and Smith, Britton D},
  journal={Monthly Notices of the Royal Astronomical Society},
  volume={393},
  number={1},
  pages={99--107},
  year={2009},
  publisher={Blackwell Publishing Ltd Oxford, UK}
}

@article{springel2003cosmological,
  title={Cosmological smoothed particle hydrodynamics simulations: a hybrid multiphase model for star formation},
  author={Springel, Volker and Hernquist, Lars},
  journal={Monthly Notices of the Royal Astronomical Society},
  volume={339},
  number={2},
  pages={289--311},
  year={2003},
  publisher={Blackwell Science Ltd Oxford, UK}
}

@article{pillepich2018simulating,
  title={Simulating galaxy formation with the IllustrisTNG model},
  author={Pillepich, Annalisa and Springel, Volker and Nelson, Dylan and Genel, Shy and Naiman, Jill and Pakmor, R{\"u}diger and Hernquist, Lars and Torrey, Paul and Vogelsberger, Mark and Weinberger, Rainer and others},
  journal={Monthly Notices of the Royal Astronomical Society},
  volume={473},
  number={3},
  pages={4077--4106},
  year={2018},
  publisher={Oxford University Press}
}

@article{hopkins2018fire,
  title={FIRE-2 simulations: physics versus numerics in galaxy formation},
  author={Hopkins, Philip F and Wetzel, Andrew and Kere{\v{s}}, Du{\v{s}}an and Faucher-Gigu{\`e}re, Claude-Andr{\'e} and Quataert, Eliot and Boylan-Kolchin, Michael and Murray, Norman and Hayward, Christopher C and Garrison-Kimmel, Shea and Hummels, Cameron and others},
  journal={Monthly Notices of the Royal Astronomical Society},
  volume={480},
  number={1},
  pages={800--863},
  year={2018},
  publisher={Oxford University Press}
}

@article{baldry2008galaxy,
  title={On the galaxy stellar mass function, the mass--metallicity relation and the implied baryonic mass function},
  author={Baldry, Ivan K and Glazebrook, Karl and Driver, Simon P},
  journal={Monthly Notices of the Royal Astronomical Society},
  volume={388},
  number={3},
  pages={945--959},
  year={2008},
  publisher={Blackwell Publishing Ltd Oxford, UK}
}

@article{akino2022hsc,
  title={HSC-XXL: Baryon budget of the 136 XXL groups and clusters},
  author={Akino, Daichi and Eckert, Dominique and Okabe, Nobuhiro and Sereno, Mauro and Umetsu, Keiichi and Oguri, Masamune and Gastaldello, Fabio and Chiu, I-Non and Ettori, Stefano and Evrard, August E and others},
  journal={Publications of the Astronomical Society of Japan},
  volume={74},
  number={1},
  pages={175--208},
  year={2022},
  publisher={Oxford University Press}
}

@article{kugel2023flamingo,
  title={FLAMINGO: calibrating large cosmological hydrodynamical simulations with machine learning},
  author={Kugel, Roi and Schaye, Joop and Schaller, Matthieu and Helly, John C and Braspenning, Joey and Elbers, Willem and Frenk, Carlos S and McCarthy, Ian G and Kwan, Juliana and Salcido, Jaime and others},
  journal={Monthly Notices of the Royal Astronomical Society},
  volume={526},
  number={4},
  pages={6103--6127},
  year={2023},
  publisher={Oxford University Press}
}

@article{vikhlinin2006chandra,
  title={Chandra sample of nearby relaxed galaxy clusters: mass, gas fraction, and mass-temperature relation},
  author={Vikhlinin, Alexey and Kravtsov, Al and Forman, W and Jones, C and Markevitch, M and Murray, SS and Van Speybroeck, L},
  journal={The Astrophysical Journal},
  volume={640},
  number={2},
  pages={691},
  year={2006},
  publisher={IOP Publishing}
}

@article{pratt2019galaxy,
  title={The galaxy cluster mass scale and its impact on cosmological constraints from the cluster population},
  author={Pratt, GW and Arnaud, M and Biviano, ANDREA and Eckert, D and Ettori, STEFANO and Nagai, D and Okabe, N and Reiprich, TH},
  journal={Space Science Reviews},
  volume={215},
  pages={1--82},
  year={2019},
  publisher={Springer}
}

@article{ettori2009cluster,
  title={The cluster gas mass fraction as a cosmological probe: a revised study},
  author={Ettori, S and Morandi, A and Tozzi, P and Balestra, I and Borgani, Stefano and Rosati, Piero and Lovisari, L and Terenziani, F},
  journal={Astronomy \& Astrophysics},
  volume={501},
  number={1},
  pages={61--73},
  year={2009},
  publisher={EDP Sciences}
}

@article{mcconnell2013revisiting,
  title={Revisiting the scaling relations of black hole masses and host galaxy properties},
  author={McConnell, Nicholas J and Ma, Chung-Pei},
  journal={The Astrophysical Journal},
  volume={764},
  number={2},
  pages={184},
  year={2013},
  publisher={IOP Publishing}
}

@article{kormendy2019coevolution,
  title={Coevolution (or not) of supermassive black holes and host galaxies: Black hole scaling relations are not biased by selection effects},
  author={Kormendy, John},
  journal={Proceedings of the International Astronomical Union},
  volume={14},
  number={S353},
  pages={186--198},
  year={2019},
  publisher={Cambridge University Press}
}

@article{davidzon2017cosmos2015,
  title={The COSMOS2015 galaxy stellar mass function-Thirteen billion years of stellar mass assembly in ten snapshots},
  author={Davidzon, I and Ilbert, O and Laigle, C and Coupon, J and McCracken, HJ and Delvecchio, IVAN and Masters, D and Capak, P and Hsieh, BC and Le F{\`e}vre, O and others},
  journal={Astronomy \& astrophysics},
  volume={605},
  pages={A70},
  year={2017},
  publisher={EDP Sciences}
}

@ARTICLE{2007A&A...464..399H,
       author = {{Hartlap}, J. and {Simon}, P. and {Schneider}, P.},
        title = "{Why your model parameter confidences might be too optimistic. Unbiased estimation of the inverse covariance matrix}",
      journal = {\aap},
         year = 2007,
        month = mar,
       volume = {464},
       number = {1},
        pages = {399-404},
          doi = {10.1051/0004-6361:20066170},
archivePrefix = {arXiv},
       eprint = {astro-ph/0608064},
 primaryClass = {astro-ph},
       adsurl = {https://ui.adsabs.harvard.edu/abs/2007A&A...464..399H}
}

@article{Nishimichi_2019,
  title="{Dark Quest. I. Fast and Accurate Emulation of Halo Clustering Statistics and Its Application to Galaxy Clustering}",
   volume={884},
   ISSN={1538-4357},
   url={http://dx.doi.org/10.3847/1538-4357/ab3719},
   DOI={10.3847/1538-4357/ab3719},
   number={1},
   journal={The Astrophysical Journal},
   publisher={American Astronomical Society},
   author = {Nishimichi, Takahiro and others},
   fullauthor={Nishimichi, Takahiro and Takada, Masahiro and Takahashi, Ryuichi and Osato, Ken and Shirasaki, Masato and Oogi, Taira and Miyatake, Hironao and Oguri, Masamune and Murata, Ryoma and Kobayashi, Yosuke and et al.},
   year={2019},
   month={Oct},
   pages={29}
}

@article{Wibking_2019,
  title="{Cosmology with galaxy–galaxy lensing on non-perturbative scales: emulation method and application to BOSS LOWZ}",
   volume={492},
   ISSN={1365-2966},
   url={http://dx.doi.org/10.1093/mnras/stz3423},
   DOI={10.1093/mnras/stz3423},
   number={2},
   journal={Monthly Notices of the Royal Astronomical Society},
   publisher={Oxford University Press (OUP)},
   author={Wibking, Benjamin D and Weinberg, David H and Salcedo, Andrés N and Wu, Hao-Yi and Singh, Sukhdeep and Rodríguez-Torres, Sergio and Garrison, Lehman H and Eisenstein, Daniel J},
   year={2019},
   month={Dec},
   pages={2872–2896}
}

@ARTICLE{kobayashi2020accurate,
       author = {{Kobayashi}, Yosuke and {Nishimichi}, Takahiro and {Takada}, Masahiro and
         {Takahashi}, Ryuichi and {Osato}, Ken},
        title = "{An accurate emulator for the redshift-space power spectrum of dark matter halos and its application to galaxy power spectrum}",
      journal = {arXiv e-prints},
         year = 2020,
        month = may,
          eid = {arXiv:2005.06122},
        pages = {arXiv:2005.06122},
archivePrefix = {arXiv},
       eprint = {2005.06122},
 primaryClass = {astro-ph.CO},
       adsurl = {https://ui.adsabs.harvard.edu/abs/2020arXiv200506122K}
}

@article{bocquet2020miratitan,
    title="{The Mira-Titan Universe. III. Emulation of the Halo Mass Function}",
    author={Sebastian Bocquet and Katrin Heitmann and Salman Habib and Earl Lawrence and Thomas Uram and Nicholas Frontiere and Adrian Pope and Hal Finkel},
    journal = {arXiv e-prints},
    year={2020},
    eprint={2003.12116},
    archivePrefix={arXiv},
    primaryClass={astro-ph.CO}
}

@article{Kwan_2013,
   title="{COSMIC EMULATION: THE CONCENTRATION-MASS RELATION FORwCDM UNIVERSES}",
   volume={768},
   ISSN={1538-4357},
   url={http://dx.doi.org/10.1088/0004-637X/768/2/123},
   DOI={10.1088/0004-637x/768/2/123},
   number={2},
   journal={The Astrophysical Journal},
   publisher={IOP Publishing},
   author={Kwan, Juliana and Bhattacharya, Suman and Heitmann, Katrin and Habib, Salman},
   year={2013},
   month={Apr},
   pages={123}
}

@article{Kwan2015,
  title={Cosmic emulation: fast predictions for the galaxy power spectrum},
  author={Kwan, Juliana and Heitmann, Katrin and Habib, Salman and Padmanabhan, Nikhil and Lawrence, Earl and Finkel, Hal and Frontiere, Nicholas and Pope, Adrian},
  journal={The Astrophysical Journal},
  volume={810},
  number={1},
  pages={35},
  year={2015},
  publisher={IOP Publishing}
}

@article{Agarwal_2014,
  title="{pkann – II. A non-linear matter power spectrum interpolator developed using artificial neural networks}",
   volume={439},
   ISSN={1365-2966},
   url={http://dx.doi.org/10.1093/mnras/stu090},
   DOI={10.1093/mnras/stu090},
   number={2},
   journal={Monthly Notices of the Royal Astronomical Society},
   publisher={Oxford University Press (OUP)},
   author={Agarwal, Shankar and Abdalla, Filipe B. and Feldman, Hume A. and Lahav, Ofer and Thomas, Shaun A.},
   year={2014},
   month={Feb},
   pages={2102–2121}
}

@article{Mead_2015,
  title="{An accurate halo model for fitting non-linear cosmological power spectra and baryonic feedback models}",
   volume={454},
   ISSN={1365-2966},
   url={http://dx.doi.org/10.1093/mnras/stv2036},
   DOI={10.1093/mnras/stv2036},
   number={2},
   journal={Monthly Notices of the Royal Astronomical Society},
   publisher={Oxford University Press (OUP)},
   author={Mead, A. J. and Peacock, J. A. and Heymans, C. and Joudaki, S. and Heavens, A. F.},
   year={2015},
   month={Oct},
   pages={1958–1975}
}

@ARTICLE{euclidemu2019,
       author = {{Knabenhans}, Mischa and others}, 
       collaboration  = "EUCLID Collaboration",
       fullauthor = {{Euclid Collaboration} and {Knabenhans}, Mischa and {Stadel}, Joachim and
         {Marelli}, Stefano and {Potter}, Doug and {Teyssier}, Romain and {Legrand
        }, Laurent and {Schneider}, Aurel and {Sudret}, Bruno and {Blot}, Linda and
         {Awan}, Saeeda and {Burigana}, Carlo and {Carvalho}, Carla Sofia and
         {Kurki-Suonio}, Hannu and {Sirri}, Gabriele},
        title = "{Euclid preparation: II. The EUCLIDEMULATOR - a tool to compute the cosmology dependence of the nonlinear matter power spectrum}",
      journal = {\mnras},
         year = 2019,
        month = apr,
       volume = {484},
       number = {4},
        pages = {5509-5529},
          doi = {10.1093/mnras/stz197},
archivePrefix = {arXiv},
       eprint = {1809.04695},
 primaryClass = {astro-ph.CO},
       adsurl = {https://ui.adsabs.harvard.edu/abs/2019MNRAS.484.5509E}
}

@article{mcdonald2017remarkable,
  title={The remarkable similarity of massive galaxy clusters from z~ 0 to z~ 1.9},
  author={McDonald, M and Allen, SW and Bayliss, M and Benson, BA and Bleem, LE and Brodwin, M and Bulbul, E and Carlstrom, JE and Forman, WR and Hlavacek-Larrondo, J and others},
  journal={The Astrophysical Journal},
  volume={843},
  number={1},
  pages={28},
  year={2017},
  publisher={IOP Publishing}
}

@misc{james_gattiker_2020_4048801,
  author       = {Gattiker, James and Klein, Natalie and Hutchings, Grant and Lawrence, Earl},
  title        = {lanl/SEPIA: v1.1},
  month        = sep,
  year         = 2020,
  publisher    = {Zenodo},
  version      = {v1.1},
  doi          = {10.5281/zenodo.4048801},
  url          = {https://doi.org/10.5281/zenodo.4048801}
}

@article{DeRose_2019,
   title="{The Aemulus Project. I. Numerical Simulations for Precision Cosmology}",
   volume={875},
   ISSN={1538-4357},
   url={http://dx.doi.org/10.3847/1538-4357/ab1085},
   DOI={10.3847/1538-4357/ab1085},
   number={1},
   journal={The Astrophysical Journal},
   publisher={American Astronomical Society},
   author={DeRose, Joseph and Wechsler, Risa H. and Tinker, Jeremy L. and Becker, Matthew R. and Mao, Yao-Yuan and McClintock, Thomas and McLaughlin, Sean and Rozo, Eduardo and Zhai, Zhongxu},
   year={2019},
   month={Apr},
   pages={69}
}

@ARTICLE{2010ApJ...715..104H,
       author = {{Heitmann}, Katrin and {White}, Martin and {Wagner}, Christian and
         {Habib}, Salman and {Higdon}, David},
        title = "{The Coyote Universe. I. Precision Determination of the Nonlinear Matter Power Spectrum}",
      journal = {\apj},
         year = 2010,
        month = may,
       volume = {715},
       number = {1},
        pages = {104-121},
          doi = {10.1088/0004-637X/715/1/104},
archivePrefix = {arXiv},
       eprint = {0812.1052},
 primaryClass = {astro-ph},
       adsurl = {https://ui.adsabs.harvard.edu/abs/2010ApJ...715..104H}
}

@ARTICLE{2009ApJ...705..156H,
       author = {{Heitmann}, Katrin and {Higdon}, David and {White}, Martin and
         {Habib}, Salman and {Williams}, Brian J. and {Lawrence}, Earl and
         {Wagner}, Christian},
        title = "{The Coyote Universe. II. Cosmological Models and Precision Emulation of the Nonlinear Matter Power Spectrum}",
      journal = {\apj},
         year = 2009,
        month = nov,
       volume = {705},
       number = {1},
        pages = {156-174},
          doi = {10.1088/0004-637X/705/1/156},
archivePrefix = {arXiv},
       eprint = {0902.0429},
 primaryClass = {astro-ph.CO},
       adsurl = {https://ui.adsabs.harvard.edu/abs/2009ApJ...705..156H}
}

@ARTICLE{2010ApJ...713.1322L,
       author = {{Lawrence}, Earl and {Heitmann}, Katrin and {White}, Martin and
         {Higdon}, David and {Wagner}, Christian and {Habib}, Salman and
         {Williams}, Brian},
        title = "{The Coyote Universe. III. Simulation Suite and Precision Emulator for the Nonlinear Matter Power Spectrum}",
      journal = {\apj},
         year = 2010,
        month = apr,
       volume = {713},
       number = {2},
        pages = {1322-1331},
          doi = {10.1088/0004-637X/713/2/1322},
archivePrefix = {arXiv},
       eprint = {0912.4490},
 primaryClass = {astro-ph.CO},
       adsurl = {https://ui.adsabs.harvard.edu/abs/2010ApJ...713.1322L}
}

@ARTICLE{2014ApJ...780..111H,
       author = {{Heitmann}, Katrin and {Lawrence}, Earl and {Kwan}, Juliana and
         {Habib}, Salman and {Higdon}, David},
        title = "{The Coyote Universe Extended: Precision Emulation of the Matter Power Spectrum}",
      journal = {\apj},
         year = 2014,
        month = jan,
       volume = {780},
       number = {1},
          eid = {111},
        pages = {111},
          doi = {10.1088/0004-637X/780/1/111},
archivePrefix = {arXiv},
       eprint = {1304.7849},
 primaryClass = {astro-ph.CO},
       adsurl = {https://ui.adsabs.harvard.edu/abs/2014ApJ...780..111H}
}

@ARTICLE{2017ApJ...847...50L,
       author = {{Lawrence}, Earl and {Heitmann}, Katrin and {Kwan}, Juliana and
         {Upadhye}, Amol and {Bingham}, Derek and {Habib}, Salman and
         {Higdon}, David and {Pope}, Adrian and {Finkel}, Hal and
         {Frontiere}, Nicholas},
        title = "{The Mira-Titan Universe. II. Matter Power Spectrum Emulation}",
      journal = {\apj},
         year = 2017,
        month = sep,
       volume = {847},
       number = {1},
          eid = {50},
        pages = {50},
          doi = {10.3847/1538-4357/aa86a9},
archivePrefix = {arXiv},
       eprint = {1705.03388},
 primaryClass = {astro-ph.CO},
       adsurl = {https://ui.adsabs.harvard.edu/abs/2017ApJ...847...50L}
}

@ARTICLE{Goodman2010,
   author = {{Goodman}, J. and {Weare}, J.},
    title = "{Ensemble samplers with affine invariance}",
  journal = {Communications in Applied Mathematics and Computational Science, Vol.~5, No.~1, p.~65-80, 2010},
     year = 2010,
   volume = 5,
    pages = {65-80},
      doi = {10.2140/camcos.2010.5.65},
   adsurl = {http://adsabs.harvard.edu/abs/2010CAMCS...5...65G}
}

@article{Heitmann2006,
  title={Cosmic calibration},
  author={Heitmann, Katrin and Higdon, David and Nakhleh, Charles and Habib, Salman},
  journal={The Astrophysical Journal},
  volume={646},
  number={1},
  pages={L1},
  year={2006},
  publisher={IOP Publishing}
}

@ARTICLE{Habib2007,
   author = {{Habib}, S. and {Heitmann}, K. and {Higdon}, D. and {Nakhleh}, C. and 
	{Williams}, B.},
    title = "{Cosmic calibration: Constraints from the matter power spectrum and the cosmic microwave background}",
  journal = {\prd},
   eprint = {astro-ph/0702348},
     year = 2007,
    month = oct,
   volume = 76,
   number = 8,
      eid = {083503},
    pages = {083503},
      doi = {10.1103/PhysRevD.76.083503},
   adsurl = {http://adsabs.harvard.edu/abs/2007PhRvD..76h3503H}
}

@article{Higdon2008,
author = {Dave Higdon and James Gattiker and Brian Williams and Maria Rightley},
title = {Computer Model Calibration Using High-Dimensional Output},
journal = {Journal of the American Statistical Association},
volume = {103},
number = {482},
pages = {570-583},
year  = {2008},
publisher = {Taylor & Francis},
doi = {10.1198/016214507000000888},
URL = {https://doi.org/10.1198/016214507000000888},
eprint = {https://doi.org/10.1198/016214507000000888}
}

@Book{Rasmussen2006,
  Title                    = {{Gaussian Processes for Machine Learning}},
  Author                   = {Rasmussen, Carl E. and Williams, Christopher},
  Publisher                = {MIT Press},
  Year                     = {2006},
  Booktitle                = {Gaussian Processes for Machine Learning},
  Url                      = {http://www.gaussianprocess.org/gpml/}
}

@MISC{numpyscipy,
   author = {{Oliphant}, T.},
    title = "{Python for Scientific Computing}",
  journal = {Computing in Science \& Engineering},
     year = 2007,
    month = May,
   volume = 9,
    pages = {10-20},
      doi = {10.1109/MCSE.2007.58}
}

@MISC{matplotlib,
   author = {{Hunter}, J.~D.},
    title = "{Matplotlib: A 2D Graphics Environment}",
  journal = {Computing in Science \& Engineering},
     year = 2007,
    month = May,
   volume = 9,
    pages = {90-95},
      doi = {10.1109/MCSE.2007.55},
}

@article{scikit,
 title={Scikit-learn: Machine Learning in {P}ython},
 author={Pedregosa, F. and others},
 fullauthor={Pedregosa, F. and Varoquaux, G. and Gramfort, A. and Michel, V.
         and Thirion, B. and Grisel, O. and Blondel, M. and Prettenhofer, P.
         and Weiss, R. and Dubourg, V. and Vanderplas, J. and Passos, A. and
         Cournapeau, D. and Brucher, M. and Perrot, M. and Duchesnay, E.},
 journal={Journal of Machine Learning Research},
 volume={12},
 pages={2825--2830},
 year={2011}
}

@article{chaikin2025colibre,
  title={COLIBRE: calibrating subgrid feedback in cosmological simulations that include a cold gas phase},
  author={Chaikin, Evgenii and Schaye, Joop and Schaller, Matthieu and Ploeckinger, Sylvia and Bah{\'e}, Yannick M and Ben{\'\i}tez-Llambay, Alejandro and Correa, Camila and Moreno, Victor J Forouhar and Frenk, Carlos S and Hu{\v{s}}ko, Filip and others},
  journal={arXiv preprint arXiv:2509.04067},
  year={2025}
}

@article{schaye2025colibre,
	author = {Schaye, Joop and Chaikin, Evgenii and Schaller, Matthieu and Ploeckinger, Sylvia and Hu{\v{s}}ko, Filip and McGibbon, Rob and Trayford, James W and Ben{\'\i}tez-Llambay, Alejandro and Correa, Camila and Frenk, Carlos S and others},
	journal = {arXiv preprint arXiv:2508.21126},
	title = {The COLIBRE project: cosmological hydrodynamical simulations of galaxy formation and evolution},
	year = {2025}}

@article{schaye2015eagle,
	author = {Schaye, Joop and Crain, Robert A and Bower, Richard G and Furlong, Michelle and Schaller, Matthieu and Theuns, Tom and Dalla Vecchia, Claudio and Frenk, Carlos S and McCarthy, IG and Helly, John C and others},
	journal = {Monthly Notices of the Royal Astronomical Society},
	number = {1},
	pages = {521--554},
	publisher = {Oxford University Press},
	title = {{The EAGLE project: simulating the evolution and assembly of galaxies and their environments}},
	volume = {446},
	year = {2015}}

@article{dave2019,
	author = {Dav{\'e}, Romeel and Angl{\'e}s-Alc{\'a}zar, Daniel and Narayanan, Desika and Li, Qi and Rafieferantsoa, Mika H and Appleby, Sarah},
	journal = {Monthly Notices of the Royal Astronomical Society},
	number = {2},
	pages = {2827--2849},
	publisher = {Oxford University Press},
	title = {SIMBA: Cosmological simulations with black hole growth and feedback},
	volume = {486},
	year = {2019}}

@article{nelson2019,
	author = {Nelson, Dylan and Springel, Volker and Pillepich, Annalisa and Rodriguez-Gomez, Vicente and Torrey, Paul and Genel, Shy and Vogelsberger, Mark and Pakmor, Ruediger and Marinacci, Federico and Weinberger, Rainer and others},
	journal = {Computational Astrophysics and Cosmology},
	number = {1},
	pages = {2},
	publisher = {Springer},
	title = {The IllustrisTNG simulations: public data release},
	volume = {6},
	year = {2019}}

@article{crain2015eagle,
  title={The EAGLE simulations of galaxy formation: calibration of subgrid physics and model variations},
  author={Crain, Robert A and Schaye, Joop and Bower, Richard G and Furlong, Michelle and Schaller, Matthieu and Theuns, Tom and Dalla Vecchia, Claudio and Frenk, Carlos S and McCarthy, Ian G and Helly, John C and others},
  journal={Monthly Notices of the Royal Astronomical Society},
  volume={450},
  number={2},
  pages={1937--1961},
  year={2015},
  publisher={Oxford University Press}
}

@article{rose2025introducing,
  title={Introducing the DREAMS project: DaRk mattEr and astrophysics with machine learning and simulations},
  author={Rose, Jonah C and Torrey, Paul and Villaescusa-Navarro, Francisco and Lisanti, Mariangela and Nguyen, Tri and Roy, Sandip and Kollmann, Kassidy E and Vogelsberger, Mark and Cyr-Racine, Francis-Yan and Medvedev, Mikhail V and others},
  journal={The Astrophysical Journal},
  volume={982},
  number={2},
  pages={68},
  year={2025},
  publisher={IOP Publishing}
}

@article{rose2025dreams,
  title={The DREAMS Project: A New Suite of 1,024 Simulations to Contextualize the Milky Way and Assess Physics Uncertainties},
  author={Rose, Jonah C and Lisanti, Mariangela and Torrey, Paul and Villaescusa-Navarro, Francisco and Garcia, Alex M and Farahi, Arya and Filion, Carrie and Brooks, Alyson M and Kallivayalil, Nitya and Kollmann, Kassidy E and others},
  journal={arXiv preprint arXiv:2512.00148},
  year={2025}
}

@article{getdist,
  title={GetDist: a Python package for analysing Monte Carlo samples},
  author={Lewis, Antony},
  journal={Journal of Cosmology and Astroparticle Physics},
  volume={2025},
  number={08},
  pages={025},
  year={2025},
  publisher={IOP Publishing}
}

@ARTICLE{emcee,
       author = {{Foreman-Mackey}, Daniel and {Hogg}, David W. and {Lang}, Dustin and
         {Goodman}, Jonathan},
        title = "{emcee: The MCMC Hammer}",
      journal = {\pasp},
         year = 2013,
        month = mar,
       volume = {125},
       number = {925},
        pages = {306},
          doi = {10.1086/670067},
archivePrefix = {arXiv},
       eprint = {1202.3665},
 primaryClass = {astro-ph.IM},
       adsurl = {https://ui.adsabs.harvard.edu/abs/2013PASP..125..306F}
}

@article{Heitmann:2013bra,
    author = "Heitmann, Katrin and Lawrence, Earl and Kwan, Juliana and Habib, Salman and Higdon, David",
    title = "{The Coyote Universe Extended: Precision Emulation of the Matter Power Spectrum}",
    eprint = "1304.7849",
    archivePrefix = "arXiv",
    primaryClass = "astro-ph.CO",
    reportNumber = "ANL-HEP-PR-13-10",
    doi = "10.1088/0004-637X/780/1/111",
    journal = "Astrophys. J.",
    volume = "780",
    pages = "111",
    year = "2014"
}
\end{document}